\documentclass{article}
\pdfoutput=1

\usepackage{jheppub}
\usepackage{amsmath}
\usepackage{bbm}
\usepackage{url}
\usepackage{color}

\definecolor{Gray}{gray}{0.5}

\def\beq{\begin{equation}}
\def\eq{\end{equation}}
\def\eeq{\end{equation}}

\def\centeron#1#2{{\setbox0=\hbox{#1}\setbox1=\hbox{#2}\ifdim
\wd1>\wd0\kern.5\wd1\kern-.5\wd0\fi
\copy0\kern-.5\wd0\kern-.5\wd1\copy1\ifdim\wd0>\wd1
\kern.5\wd0\kern-.5\wd1\fi}}
\def\ltap{\;\centeron{\raise.35ex\hbox{$<$}}{\lower.65ex\hbox{$\sim$}}\;}
\def\gtap{\;\centeron{\raise.35ex\hbox{$>$}}{\lower.65ex\hbox{$\sim$}}\;}
\def\gsim{\mathrel{\gtap}}

\def\chii0{\chi_i^0}
\def\chij0{\chi_j^0}

\def\foursqr#1#2{{\vcenter{\vbox{
 \hrule height.#2pt
 \hbox{\vrule width.#2pt height#1pt \kern#1pt
 \vrule width.#2pt}
 \hrule height.#2pt
 \hrule height.#2pt
 \hbox{\vrule width.#2pt height#1pt \kern#1pt
 \vrule width.#2pt}
 \hrule height.#2pt
     \hrule height.#2pt
 \hbox{\vrule width.#2pt height#1pt \kern#1pt
 \vrule width.#2pt}
 \hrule height.#2pt
     \hrule height.#2pt
 \hbox{\vrule width.#2pt height#1pt \kern#1pt
 \vrule width.#2pt}
 \hrule height.#2pt}}}}
\def\psqr#1#2{{\vcenter{\vbox{\hrule height.#2pt
 \hbox{\vrule width.#2pt height#1pt \kern#1pt
 \vrule width.#2pt}
 \hrule height.#2pt \hrule height.#2pt
 \hbox{\vrule width.#2pt height#1pt \kern#1pt
 \vrule width.#2pt}
 \hrule height.#2pt}}}}
\def\sqr#1#2{{\vcenter{\vbox{\hrule height.#2pt
 \hbox{\vrule width.#2pt height#1pt \kern#1pt
 \vrule width.#2pt}
 \hrule height.#2pt}}}}

\def\figin{\epsfcheck\figin}\def\figins{\epsfcheck\figins}
\def\epsfcheck{\ifx\epsfbox\UnDeFiNeD
\message{(NO epsf.tex, FIGURES WILL BE IGNORED)}
\gdef\figin##1{\vskip2in}\gdef\figins##1{\hskip.5in}
\else\message{(FIGURES WILL BE INCLUDED)}%
\gdef\figin##1{##1}\gdef\figins##1{##1}\fi}
\def\DefWarn#1{}
\def\figinsert{\goodbreak\midinsert}
\def\ifig#1#2#3{\DefWarn#1\xdef#1{fig.~\the\figno}
\writedef{#1\leftbracket fig.\noexpand~\the\figno}%
\figinsert\figin{\centerline{#3}}\medskip\centerline{\vbox{\baselineskip12pt
\advance\hsize by -1truein\noindent\footnotefont{\bf
Fig.~\the\figno:\ } \it#2}}
\bigskip\endinsert\global\advance\figno by1}

\def\fig#1#2#3#4{\vskip 0.5cm \begingroup \midinsert \centerline{
\psfig{file=#1,width=#2}} \vskip 0.4cm
\global\advance\figno by 1
\centerline{\vbox{\baselineskip=12pt \noindent Figure \the\figno: #3}}
\endinsert \endgroup {\xdef#4{\the\figno}} }

\def\figcrop#1#2#3#4#5#6#7#8{\vskip 0.5cm \begingroup \midinsert \centerline{
\psfig{file=#1,width=#2,bbllx=#3,bblly=#4,bburx=#5,bbury=#6}} \vskip 0.4cm
\global\advance\figno by 1
\centerline{\vbox{\baselineskip=12pt \noindent Figure \the\figno: #7}}
\endinsert \endgroup {\xdef#8{\the\figno}} \vskip .5cm}

\def\figlabel#1{\xdef#1{\the\figno}}
\def\encadremath#1{\vbox{\hrule\hbox{\vrule\kern8pt\vbox{\kern8pt
\hbox{$\displaystyle #1$}\kern8pt}
\kern8pt\vrule}\hrule}}
\def\underarrow#1{\vbox{\ialign{##\crcr$\hfil\displaystyle
 {#1}\hfil$\crcr\noalign{\kern1pt\nointerlineskip}$\longrightarrow$\crcr}}}

\def\sigmabr{\sigma \!  \cdot \!  {\rm Br}}
\def\sigmabrr{\sigma \! \! \cdot \!  \! {\rm Br}}

\title{Exclusive Signals of an Extended Higgs Sector
}

\author[a,b]{Nathaniel Craig,}

\author[a]{Scott Thomas}

\affiliation[a]{Department of Physics, Rutgers University \\
Piscataway, NJ 08854 }
\affiliation[b]{ School of Natural Sciences, Institute for Advanced Study \\
Princeton, NJ 08540}

\preprint{RU-NHETC-2012-15}

\abstract{
Expectations for the magnitude of Higgs boson signals in 
standard Higgs search channels at the LHC
relative to Standard Model (SM) expectations 
are investigated within the framework of various types of 
CP and flavor conserving two Higgs doublet models (2HDMs). 
Signals of the SM-like Higgs boson in different classes of 2HDM
may be parameterized in terms of particular two-dimensional sub-spaces
of the general four-dimensional space of Higgs couplings to 
the massive vector bosons, top quark, bottom quark, and tau lepton.  
We find fairly strong correlations among the 
inclusive di-photon channel and the 
exclusive di-photon and di-tau channels from 
vector boson fusion or associated production. 
Order one deviations 
from SM expectations in some of these channels could 
provide discriminating power among various types of 2HDMs.  
The ratio of exclusive di-photon to di-tau channels is particularly 
sensitive to deviations from SM expectations.  
We also emphasize that deviations from SM expectations in 
standard Higgs search channels may imply observable signals 
of non-SM-like Higgs bosons in some of these same channels, 
in particular in di-photon and di-vector boson channels.   
The results cataloged here provide a roadmap for interpreting 
standard Higgs search channels in the context of 2HDMs.  
}

\begin{document}

\maketitle

\section{Introduction}

The Large Hadron Collider (LHC) is in the process of revealing the mechanism of electroweak symmetry breaking, with the discovery of a Higgs-like scalar near
125 GeV \cite{CMS:2012gu, ATLAS:2012gk}.  Given the discovery of a new scalar in conventional Higgs search channels, the immediate question is whether its properties 
are those of the Standard Model Higgs boson. 
This question will begin to be probed in several ways with a relatively low amount of integrated luminosity, both by measuring 
cross section times branching ratios in standard Higgs search channels, and 
by possibly observing additional degrees of freedom 
in these or other channels.  
Unsurprisingly, considerable effort has recently been devoted to studying the 
potential implications of standard search channels for probing  
the couplings of a Higgs near 125 GeV \cite{fits} and the prospects for measuring couplings with additional integrated luminosity \cite{Klute, Wells}.\footnote{Sharper questions, such as whether the vertex structure of the coupling between the Higgs and gauge bosons corresponds to that expected from electroweak symmetry breaking, may also be answerable with greater integrated luminosity \cite{Plehn:2001nj}.} 

Although it is possible to envision special-purpose measurements
aimed specifically at measuring the couplings of a Standard Model-like (SM-like) Higgs boson, or
new searches aimed at discovering additional states, 
a great deal may be learned simply by studying the effects of non-standard Higgs properties on ongoing searches in the standard Higgs search channels. 
At present, considerable experimental effort has been devoted to these standard Higgs 
search channels -- i.e., 
the ones most promising for observing production and decay of the Standard Model Higgs boson. 
The dominant production mode in this case 
is through gluon-gluon fusion ($gg$F), $gg \to h$, which benefits from a large cross section and the resonant production of the Higgs. 
But there are also a variety of ancillary channels in which the Higgs is produced in association with other quarks or vector bosons; these channels typically have smaller backgrounds due to the availability of additional reconstructable objects in the final state. 
These are, in order of decreasing production rate: weak vector boson fusion (VBF), $qq \to qqh$; $Vh$ associated production, $q\bar q' \to Wh, Zh$; and $t \bar t h$ associated production, $q\bar q, gg \to t \bar t h$. For a given Higgs final state, it is possible to measure both {\it inclusive} production, in which all production channels are combined; and {\it exclusive} production, in which the distinctive properties of certain associated prioduction 
channels may be isolated. While the former offers the largest possible signal, the latter offers lower backgrounds and the possibility of distinguishing the magnitude 
of various Higgs couplings. 

On the decay side, the final states with the greatest sensitivity are $h \to \gamma \gamma$, $h \to ZZ^* \to 4 \ell$, and $h \to WW^* \to \ell \ell \nu \nu$. The distinctive final state topology of the di-photon channel makes it a crucial search channel for lighter masses despite the relatively small signal. Similarly, the cleanliness of the $2 \ell$ and $4 \ell$ final states from $WW^*$ and $ZZ^*$ production make $h \to \ell \ell \nu \nu$ and $h \to 4 \ell$ particularly attractive. The $h \to \tau \tau$ and $h \to b \bar b$ final states are also fairly promising due to the potentially large branching ratios, though the size of QCD backgrounds to these processes means that they are plausibly observable only in associated production channels, where the additional tagging information helps to improve the signal-to-background ratio. A final possibility is to probe the various non-resonant multilepton final states available through both gluon-gluon fusion and various associated production modes \cite{multilepton}. While this approach does not offer sharp mass resolution, the low backgrounds to many nonresonant same-sign two-, 
three- and four-lepton Higgs final states provides considerable sensitivity.

Searches for additional scalars 
in many of these standard Higgs channels should continue after discovery, 
and it is worth examining how these 
probe for deviations from a Standard Model Higgs sector. 
Since the production and decay rates of the Standard Model Higgs are completely determined in 
terms of its mass, 
it is possible to compare cross section times branching ratio 
signals in standard Higgs channels to the SM expectation, 
and thereby obtain information about the relative strength of the couplings
of a SM-like Higgs boson.   
In this respect, exclusive standard channels are particularly useful.\footnote{For recent related work see e.g. \cite{exclusive}.} 
The background to many exclusive channels is low, putting a variety of channels within reach at low luminosity. Different exclusive production channels depend on different Higgs couplings, and therefore may be used to separately probe Higgs couplings to fermions and gauge bosons. 
Perhaps most attractively, it is possible to measure the ratios of various exclusive channels, in which case many systematic errors drop out and functions of ratios of 
couplings may be measured with the greatest possible accuracy. Needless to say, these channels may also be sensitive to additional states in the Higgs sector and can play a further role in their discovery and characterization.

In this work we examine the range of signals appearing in standard Higgs search 
channels in minimal extensions of the electroweak symmetry breaking sector. We focus on perhaps the simplest class of extended electroweak symmetry breaking: theories with an additional Higgs doublet. Such two Higgs doublet models (2HDM) are well-motivated on their own by beyond-the-Standard-Model theories such as supersymmetry, and also serve as comprehensive effective theories for more exotic means of electroweak symmetry breaking. As simple effective theories for extended EWSB, 2HDM have attracted considerable attention with regard to Higgs searches at the LHC; for recent related work on 2HDM signals see e.g.  \cite{2hdm}.

We focus on standard Higgs search 
channels that should 
be accessible with a relatively low amount of integrated luminosity, 
roughly ${\cal O}(30-50)$ fb$^{-1}$ at each LHC experiment; 
this amounts to data adequate for observation of various exclusive production and decay channels. 
We neglect other channels that are more model-dependent, 
including additional non-Standard Model 
decays among the various scalars of a 2HDM. 
For the sake of definiteness, we will assume when necessary
that 
the mass of the SM-like Higgs boson, $h$, is $m_h = 125$ GeV, with 
other scalars heavier. 
Our focus lies on two particularly interesting phenomena: 
(1) the range and correlation of (ratios of) cross section times branching ratio measurements 
for the SM-like Higgs boson 
in both inclusive and exclusive channels that can be realized
in various 2HDM;  and 
(2) the relation between potential 
discrepancies in these measurements 
and the contribution of additional Higgs bosons to standard Higgs search channels. 
Our purpose is not to quantify the precision with which
measurements may be performed 
in various channels, but rather to explore possible ranges for (sizable) deviations from Standard Model Higgs predictions that could arise in the 
post discovery-level data set. 
In this sense we provide a road map for the interpretation of any deviations that might 
persist in early observations of both inclusive and 
exclusive channels  within the framework of two Higgs doublets.

Given discovery of a Higgs-like scalar at 125 GeV, it is already possible to investigate how closely its properties resemble those of a Standard Model Higgs. 
Numerous fits of Higgs boson couplings 
to individual particles using recent experimental results 
have been performed at both the theory level \cite{fits} and by the experimental collaborations themselves \cite{CMS:2012gu, ATLAS:2012gk}. 
At present, the signal fits are governed by a small number of channels and the errors on these fits remain large due to the limited statistics
 of the discovery-level data set, particularly for exclusive channels with smaller cross sections times branching ratios or challenging backgrounds. 
Fits to Higgs boson couplings may be interpreted in terms of our 2HDM roadmap to probe 
 the range of deviations in exclusive Higgs channels that have yet to be 
 observed or measured with precision.

 To this end, in Section \ref{stanchan} we enumerate the various inclusive and exclusive standard Higgs channels that should 
 be observable with relatively low integrated luminosity. 
 In Section \ref{2hdm} we review the structure of the simplest 2HDMs 
 and their couplings to Standard Model fermions and gauge bosons. 
 In Section \ref{inclusive} we consider the ranges of various 
 (ratios of) inclusive cross section time branchings 
 of the SM-like Higgs $h$ in various 2HDMs. 
 We then turn in Section \ref{exclusive} to consider (ratios of) exclusive cross section 
 times branching ratios of the SM-like Higgs $h$, 
as well as correlations among various exclusive channels. 
In Section \ref{heavy} we turn our attention to the other non-SM-like  
scalars in 2HDMs. 
We show that within 2HDMs, discrepancies of certain 
cross section times branching ratios for the SM-like Higgs 
can imply 
that some of the heavier scalars may be observed in standard Higgs search 
channels. 
In Section \ref{fits} we briefly consider the implications for 2HDM 
of current experimental fits to  inclusive signals of the recently-discovered resonance.
Future directions are discussed in Section \ref{conclusion}.

\section{Standard Higgs channels}
\label{stanchan}

As discussed in the introduction, there are a number of standard channels, both inclusive and exclusive, 
that should be observable with relatively low integrated luminosity
after the discovery of a SM-like Higgs boson. 
The exclusive channels potentially available in the early 
post-discovery data set are particularly attractive. 
Although the cross sections for exclusive channels are considerably smaller than their inclusive counterparts, the backgrounds are typically lower. The ability to isolate a particular production process 
makes these exclusive channels 
particularly useful, as does the possibility of measuring exclusive ratios, 
for which many systematics drop out. In this sense, exclusive ratios 
may provide the first sensitive probe of Higgs couplings.

To be more specific, in Table \ref{tab:channels} we list the standard Higgs channels we estimate 
to be potentially observable in the early data set, assuming Standard Model 
cross sections for a Higgs mass of $m_h = 125$ GeV. 
The di-photon final state is the most promising, having been observed in both inclusive production and VBF, and should be observable in every production channel. 
The $t \bar t h$ di-photon signal is very small, 
but nonetheless should be observable above the (relatively low) background.  
The $h \to WW^* \to 2 \ell 2 \nu$ final state has been observed in inclusive production and should be separately observable in VBF. It is also possible that the leptonic $WW$ decays will be accessible in $Vh$ and potentially $t \bar t h$ associated production through the nonresonant multi-lepton final states. Similarly, $h \to ZZ^* \to 4 \ell$ is already visible in inclusive production and should be separately visible in VBF. The prospects for observing leptonic $h \to ZZ^*$ decays in $Vh$ and $t \bar t h$ are somewhat poor, due to the large Standard Model background for multi-lepton final states involving two or more $Z$ bosons. The decay $h \to \tau \tau$ should be visible in the exclusive VBF channel when one or both $\tau$'s decay leptonically; 
backgrounds for $h \to \tau \tau$ produced in gluon fusion are prohibitively high, while the production cross sections for other associated channels are low. Finally, the purely hadronic decay $h \to \tau \tau$ should be observable in the $Vh$ and VBF associated production channels. 


\begin{table}[h]
\caption{Higgs search channels that are 
potentially observable with relatively low integrated luminosity
post-discovery of a light SM-like Higgs boson.
Gray checks ($tth$ with $h \to \gamma \gamma$ or 
$h \to WW^*$; 
$Vh$ with $h \to WW^*$; 
VBF with $h \to b \bar{b}$) 
denote borderline channels that may be more challenging to observe in early data.}
\begin{center}
\begin{tabular}{|c|cccc|} \hline
& Inc. & VBF & $Vh$ & $t \bar t h$ \\ \hline
$\gamma \gamma$ & $\checkmark $ & $\checkmark $ & $\checkmark $ & $\textcolor{Gray}{\checkmark }$  \\
$WW^*$ &$\checkmark $ & $\checkmark $ & $\textcolor{Gray}{\checkmark }$ & $\textcolor{Gray}{\checkmark }$  \\
$ZZ^*$&$\checkmark $ & $\checkmark $ & $-$ & $-$  \\
$\tau \tau$ &$\textcolor{Gray}{\checkmark }$ & $\checkmark $ & $-$ & $-$   \\
$b \bar b$ &$-$ & $\textcolor{Gray}{\checkmark }$ & $\checkmark $ & $-$ \\
\hline
\end{tabular}
\end{center}
\label{tab:channels}
\end{table}%

One of the most natural questions to ask at this stage is  to what accuracy the Higgs couplings may be extracted with a low luminosity data set.  There have been several theory-level 
studies to this end in the past 
decade \cite{Zeppenfeld:2000td, Duhrssen:685538, Duhrssen:2004cv, Ruwiedel:2007zz, Lafaye:2009vr, Klute}. 
Such theory studies must be treated with caution, since a detailed treatment of the actual 
systematic errors present in measurements of 
cross section times branching ratios, particularly  
in (ratios of) exclusive channels, is necessarily an experiment-level question.
Nonetheless, they provide some rough
guide of the approximate accuracy with which couplings and ratios 
might be measured at a given integrated luminosity. 
Of these,  \cite{Klute} is the most germane to the situation at hand, investigating the accuracy with which couplings and ratios of exclusive processes might be measured for a 125 GeV Higgs using 7.5 - 17.5 fb$^{-1}$ of integrated luminosity at 7-8 TeV and 30 fb$^{-1}$ of integrated luminosity at 14 TeV. The one sigma error bars for any given channel at 7-8 TeV with 17.5 fb$^{-1}$ are larger than $\pm 20 \%$ of the Standard Model value, with the $hWW$ coupling measurable to within $\sim \pm 25 \%$; the uncertainties for $htt$ (hence also $hgg$), $hbb$, and $h \tau \tau$ couplings are $\mathcal{O}(1)$ fractions of the SM value. The prospects for exclusive ratios are similar, with the most promising ratio being that of $g_{h ZZ} / g_{h WW}$.  The $g_{h \tau \tau} / g_{h WW}$ should also be promising \cite{Lafaye:2009vr}, measurable via the ratio of VBF exclusive production for the $h \to WW \to 2 \ell 2 \nu$ and $h \to \tau \tau$ final states. 
These projections provide a useful guide to the channels that may most strongly constrain deviations from SM couplings at low luminosity, but further detailed study is beyond the scope of the current work.

\section{Extended Higgs sectors}
\label{2hdm}

In principle there are many possible extensions and deformations of the minimal 
Higgs sector. 
Here we restrict our focus to perhaps the simplest class of extensions: theories with two Higgs fields transforming as doublets under $SU(2)_L$ with unit $U(1)_Y$ charge. Such 2HDMs provide a general effective theory framework for extensions of the electroweak symmetry breaking sector, supersymmetric or otherwise. Of the eight real scalars present in a  
2HDM, three are eaten by electroweak symmetry breaking, 
leaving five physical scalars: with CP conservation, 
the CP even neutral Higgses $h$ and $H$; the CP odd pseudoscalar $A$; and the charged Higgses $H^\pm$.

 Given more than one Higgs doublet, a sufficient condition for 
 the absence of tree-level FCNCs is guaranteed 
by the Glashow-Weinberg condition \cite{Glashow:1976nt}
that all fermions of a given gauge representation  
receive their mass through renormalizable 
couplings to precisely one Higgs doublet. 
In the case of two Higgs doublets (denoted $\Phi_1$ and $\Phi_2$), 
the Glashow-Weinberg condition 
is satisfied by precisely four discrete 
types of 2HDM distinguished by the possible assignments of 
fermion couplings. 
By convention $\Phi_2$ is fixed to be the Higgs doublet that couples to $Q \bar u$. This leaves four possible choices of couplings to $Q \bar d$ and $L \bar e$. 
Of these four models, Type I 
with all fermions coupled to one doublet contains the fermi-phobic Higgs as a limit
(in which the SM-like Higgs is isolated in the other doublet).  
 Type II is MSSM-like, in that this is the only choice of charge assignments consistent with a holomorphic superpotential.  What we choose to call 
Type III is also known as lepton-specific, since it assigns a Higgs doublet solely to
leptons.\footnote{To avoid confusion, we emphasize that this is distinct from the general 
2HDM with flavor-violating couplings, often referred to in the literature as Type III. 
There also appears to be no universal convention relating the III and IV labels to lepton-specific and flipped 2HDM.}  What we choose to call Type IV is also known as flipped, for the obvious reason that the leptons have a flipped coupling relative to Type II. These possible couplings are illustrated in Table \ref{tab:2hdm}; for a comprehensive review, see \cite{Branco:2011iw}.

\begin{table}[htdp]
\caption{Higgs boson couplings to $SU(2)_L$ singlet fermions in the four discrete types 
of 2HDM models 
that satisfy the Glashow-Weinberg condition.}
\begin{center}
\begin{tabular}{|c|c|c|c|c|} \hline
& 2HDM I & 2HDM II & 2HDM III & 2HDM IV \\ \hline
$u$ & $\Phi_2$ & $\Phi_2$ & $\Phi_2$ & $\Phi_2$ \\
$d$ & $\Phi_2$ & $\Phi_1$ & $\Phi_2$ & $\Phi_1$ \\
$e$ & $\Phi_2$ & $\Phi_1$ & $\Phi_1$ & $\Phi_2$  \\ \hline
\end{tabular}
\end{center}
\label{tab:2hdm}
\end{table}%

If we restrict ourselves to extended EWSB sectors with  two Higgs doublets 
that satisfy the Glashow-Weinberg condition 
with no tree-level FCNCs, and CP conservation, then the 
renormalizable tree-level couplings of all five scalar degrees of freedom to Standard Model fermions and massive gauge bosons are fixed in terms of two parameters: the mixing angle $\alpha$ of the two CP even neutral mass eigenstates $h, H$; and the angle $\beta$, which parameterizes the relative contribution of each doublet to EWSB via $\tan \beta \equiv \langle \Phi_2 \rangle / \langle \Phi_1 \rangle.$ 
In particular, this involves no additional assumptions about the form of the full non-renormalizable 
scalar potential (beyond CP conservation). 
This is to be contrasted with (CP and flavor conserving) 
multi-Higgs theories 
for which the tree-level couplings to up- and down-type quarks, leptons, and massive 
gauge bosons are in general all independent.  
The general tree-level couplings of  
CP-conserving 2HDMs (with the Glashow-Weinberg condition) are therefore restricted to particular 
two-dimensional sub-spaces of the general four-dimensional space of Higgs couplings 
to the Standard Model fermions and massive gauge boson.  
This is also to be contrasted with the single Higgs 
theory with general non-renormalizable 
couplings in which the coupling to every Standard Model state is 
independent (with deviations from renormalizable couplings parameterized 
by non-renormalizable operators).

The two-parameter scalings of the neutral Higgs boson couplings 
to SM fermions and massive gauge bosons, 
relative to the SM values, 
for the four types of 2HDMs 
are shown in Table \ref{tab:couplings}. 
This implies that partial widths for tree-level decays of the Higgs scalars to Standard Model final states also depend only on $\alpha, \beta$ relative to SM Higgs partial widths. These relations are modified at the quantum level, but in the framework here such corrections are perturbatively small.\footnote{For recent studies of cases (such as certain limits of the MSSM) in which additional degrees of freedom beyond the 2HDM may significantly alter partial widths at one loop, see e.g.  \cite{loops}. } Similarly, note that the loop-induced decay 
$h \to \gamma \gamma$ exhibits a very mild dependence on additional parameters, since the charged Higgs $H^\pm$ may run in the loop along with Standard Model fields. However, unless the charged Higgses are particularly light, this effect is negligible compared to contributions from $W$ and top loops, being suppressed relative to $W$ and top loops by a factor of 
${\cal O}(m_W^2 / m_{H^\pm}^2)$. 
As such, in what follows we will neglect corrections from $H^\pm$ to the $h \to \gamma \gamma$ rate.

Of course, the actual branching ratios of Higgs scalars to SM final states depend on the sum of their partial widths, which may include appreciable decays to other Higgs scalars that are sensitive to the details of the scalar potential. Nonetheless, if $h$ is the lightest Higgs, then it has the same decay channels as the SM Higgs, with no additional decay modes. It also has the same production channels as the SM Higgs, though there may be additional contributions to production modes via heavy scalars; the size of these contributions depends on the masses and branching ratios of the heavy scalars. If these additional contributions are negligible, the observable production cross sections and branching ratios in standard Higgs channels of the scalar $h$ are completely determined relative to the SM Higgs by the parameters $\alpha$ and $\beta$. In contrast, the branching ratios of the heavier scalar states depend on the mass orderings; we will reserve a discussion of these decays for Section 6.

\begin{table}[h]
\caption{Tree-level couplings of the neutral Higgs bosons
to up- and down-type quarks, leptons, and massive gauge bosons 
in the four types of 2HDM models relative to the SM Higgs boson couplings as a function of $\alpha$ and $\beta$.}
\begin{center}
\begin{tabular}{|c|c|c|c|c|} \hline
& 2HDM I & 2HDM II & 2HDM III & 2HDM IV \\ \hline
$hVV$ & $\sin(\beta - \alpha)$ &  $\sin(\beta - \alpha)$ &  $\sin(\beta - \alpha)$ &  $\sin(\beta - \alpha)$ \\
$h Q u $ & ${\cos \alpha}/{\sin \beta}$ & ${\cos \alpha}/{\sin \beta}$ & ${\cos \alpha}/{\sin \beta}$& ${\cos \alpha}/{\sin \beta}$  \\
$h Q d$ & ${\cos \alpha}/{\sin \beta}$ & $- {\sin \alpha}/{\cos \beta}$ & ${\cos \alpha}/{\sin \beta}$& $- {\sin \alpha}/{\cos \beta}$  \\
$h L e$ & ${\cos \alpha}/{\sin \beta}$ & $- {\sin \alpha}/{\cos \beta}$ & $- {\sin \alpha}/{\cos \beta}$& ${\cos \alpha}/{\sin \beta}$  \\ \hline
$HVV$ & $\cos(\beta - \alpha)$ & $\cos(\beta - \alpha)$ & $\cos(\beta - \alpha)$& $\cos(\beta - \alpha)$   \\
$H Q u$ & ${\sin \alpha}/{\sin \beta}$ & ${\sin \alpha}/{\sin \beta}$ & ${\sin \alpha}/{\sin \beta}$& ${\sin \alpha}/{\sin \beta}$  \\
$H Q d$ & ${\sin \alpha}/{\sin \beta}$ & ${\cos \alpha}/{\cos \beta}$ & ${\sin \alpha}/{\sin \beta}$&${\cos \alpha}/{\cos \beta}$  \\
$H L e$ & ${\sin \alpha}/{\sin \beta}$ & ${\cos \alpha}/{\cos \beta}$ &${\cos \alpha}/{\cos \beta}$& ${\sin \alpha}/{\sin \beta}$  \\\hline
$AVV$ & 0 & 0 & 0 & 0 \\
$AQu$ & $\cot \beta$ & $\cot \beta$ & $\cot \beta$& $\cot \beta$   \\
$AQd$ & $-\cot \beta$ & $\tan \beta$& $- \cot \beta$ & $\tan \beta$   \\ 
$ALe$ & $- \cot \beta$ & $\tan \beta$  & $\tan \beta$& $- \cot \beta$  \\ \hline
\end{tabular}
\end{center}
\label{tab:couplings}
\end{table}%

Assuming the SM-like Higgs, $h$, 
is the lightest scalar, we may efficiently estimate the NLO production 
cross section and branching ratios for the light CP even neutral Higgs $h$ in a 2HDM from known NLO rates for a Standard Model Higgs of the same mass by applying the LO coupling ratios as a function of $\alpha, \beta$. For example, we may estimate NLO branching ratios via
$$
\Gamma_{\rm NLO}(X \to Y) \simeq \Gamma^{SM}_{\rm NLO} (X \to Y) ~ 
\frac{\Gamma_{\rm LO}(X \to Y)}{\Gamma^{SM}_{\rm LO}(X \to Y)}
$$
where the LO ratio $ {\Gamma_{\rm LO}(X \to Y)}/{\Gamma^{SM}_{\rm LO}(X \to Y)}$ may be obtained by applying the parametric scalings in Table \ref{tab:couplings} to the relevant Standard Model processes. This procedure is straightforward for tree-level couplings, and may be extended to the leading-order loop-level couplings assuming only SM particles run in the loops \cite{Gunion:1989we}. Likewise, we may estimate the NLO production cross section for a given process via
$$ 
\sigma_{\rm NLO}(X \to Y) \simeq \sigma^{SM}_{\rm NLO}(X \to Y) ~ 
 \frac{\Gamma_{\rm LO}(X \to Y)}{\Gamma^{SM}_{\rm LO}(X \to Y)}.
$$
We obtain the cross sections for each SM Higgs boson production channel 
and  branching ratios for 
for a Higgs mass of 125 GeV 
from the LHC Higgs Cross Section Group \cite{LHCHiggsCrossSectionWorkingGroup:2011ti}.  
In what follows, the cross sections for various channels are important largely for determining their relative weight in inclusive processes. For simplicity we focus on cross sections at 7 TeV, and note that the relative contributions of various channels do not change significantly between 7 and 8 TeV.  

\begin{figure}[h] 
   \centering
   \includegraphics[width=3in]{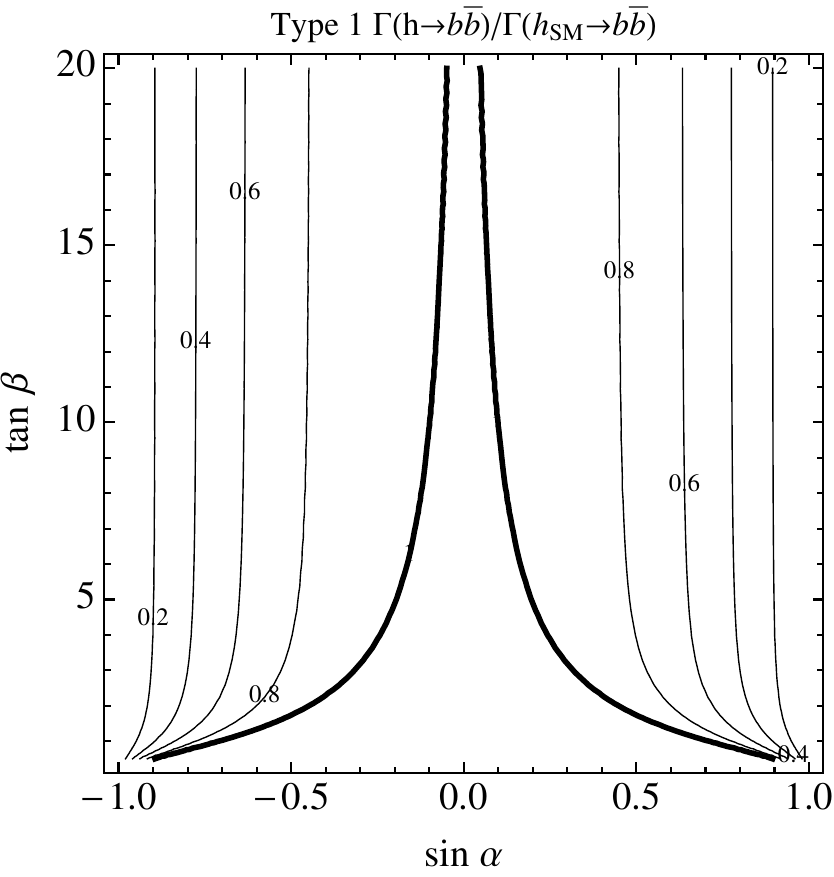} 
      \includegraphics[width=3in]{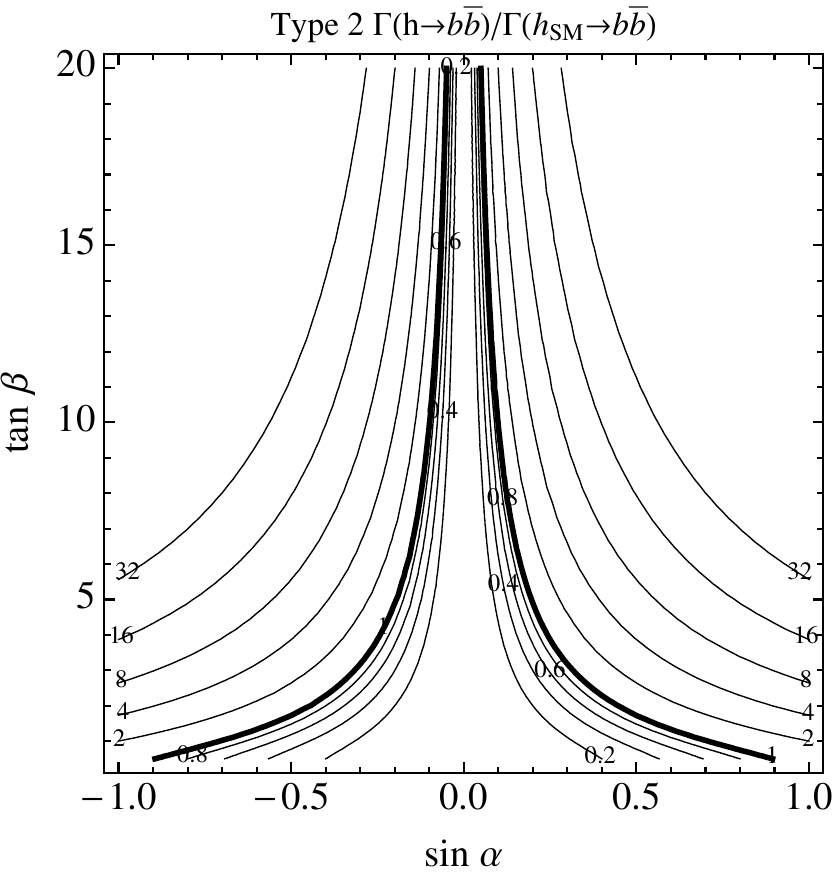}
   \caption{Contours of 
   $\Gamma(h \to b \bar b) / \Gamma(h_{SM} \to b \bar b)$ 
   for the SM-like Higgs boson as a function of $\sin \alpha$ and 
   $\tan \beta$ in Type 1 2HDM (left) and Type 2 2HDM (right). 
   The Type 3 model is parametrically similar to Type 1, while Type 4 is similar to Type 2. 
   Thick black lines denote the SM value.  
   }
   \label{fig:widthbb}
\end{figure}

\begin{figure}[h] 
   \centering
   \includegraphics[width=3in]{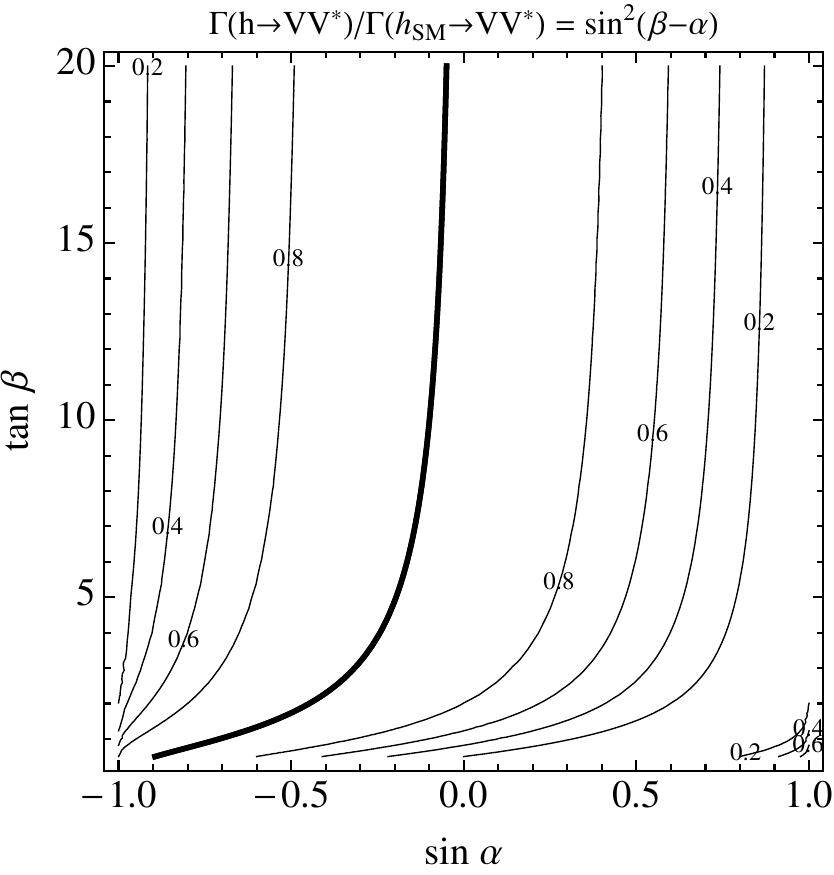} 
   \caption{
   Contours of 
   $\Gamma(h \to VV^{(*)}) / \Gamma(h_{SM} \to VV^{(*)}) = \sin^2(\beta - \alpha)$ 
   for the SM-like Higgs boson as a function of $\sin \alpha$ and 
   $\tan \beta$ in any of the 2HDMs.  
   Thick black lines denote the SM value corresponding to the 
   $\sin^2(\beta - \alpha)=1$ 
   alignment 
   limit.   
 }
   \label{fig:width}
\end{figure}

There are two key quantities that control much of the parametric behavior in various Higgs channels: the coupling of the Higgs to $W$ and $Z$ bosons, and the partial width $\Gamma(h \to b \bar{b})$. 
The VBF and $Vh$ exclusive production $\sigma \! \cdot \! {\rm Br}$ 
scale as $g_{hVV}^2 \propto \sin^2(\beta - \alpha)$, as does the partial 
width $\Gamma(h \to VV^*)$. 
As $\sin^2(\beta - \alpha) \to 1$, the vector 
boson production and decay modes of $h$ approach those of the Standard Model Higgs, 
while the vector boson production and decay of the heavy neutral scalar $H$ approach zero. 
Thus $\sin^2(\beta - \alpha)$ -- which a is universal 
scaling of the $hVV^*$ partial width among the various 2HDM 
types -- underlies much of the parametric behavior of key production and decay modes. 
The condition $\sin^2(\beta - \alpha)=1$ is obtained when the SM-like Higgs, $h$, is 
aligned with the two Higgs vacuum expectation values.  
We correspondingly designate $\sin^2(\beta - \alpha) \to 1$ 
as an alignment limit. 
The alignment limit may be obtained in certain decoupling limits in which the 
masses of the non-SM-like Higgs states become large, but may 
also be considered more generally as a limit of the couplings 
with a fixed Higgs boson mass spectrum. 
A second useful quantity is the partial width $\Gamma(h \to b \bar b)$, insofar as it is 
typically the dominant decay mode and controls the total width of the Higgs 
(assuming no significant decay products beyond the standard channels). 
Since the $\sigma \! \cdot \! {\rm Br}$ for many standard Higgs channels is 
inversely proportional to the total width, the parametric behavior of the 
dominant $\Gamma(h \to b \bar b)$ partial width typically governs the 
branching ratio for rare decays. These two quantities are 
shown in Figs.~\ref{fig:widthbb} and \ref{fig:width}.

The parametric behavior of these quantities as a function of $\alpha$ and $\beta$ often suffices to understand the scaling of various exclusive channels. In particular, this implies that the signals of Type 1 and Type 3 2HDM are similar to each other in most standard channels (particularly for di-photon and diboson final states, for both inclusive and exclusive production); the same is likewise true for Type 2 and Type 4 2HDM. These similarities arise because in each case the quark couplings are identical for the pairs of 2HDM, so in particular the scaling of the $h \to b \bar{b}$ partial widths (as well as the $h t \bar t$ couplings that governs the gluon fusion production rate) are identical. The only substantial distinction arises in standard channels with ditau final states, since the lepton couplings differ.

\section{SM-like Higgs inclusive production}
\label{inclusive}

Inclusive production channels provided the first decisive evidence for the Higgs, in large part due to the size of the $gg$ fusion production cross section. However, not all final states of the Higgs may be efficiently probed through inclusive production; high-background channels such as $h \to \tau \tau$ and $h \to b \bar b$ require additional discriminants such as forward jets or associated leptons in order to efficiently distinguish signal from background. Of the possible inclusive processes, the  $h \to \gamma \gamma$ and various  $h \to VV^*$ channels are the most promising probes for deviations from SM couplings, particularly $h \to WW^* \to 2 \ell 2 \nu$ and $h \to ZZ^* \to 4 \ell$. These latter states all have identical parametric scaling in the 2HDM under consideration, so we may treat them collectively.

We approximate the inclusive production $\sigma \! \cdot \! {\rm Br}$ by summing over the production cross sections of various channels. This procedure does not account for possible differences in the experimental acceptance between various production channels. However, the corrections in this case due to unknown acceptance are at the very most $\mathcal{O}(\sigma^{SM}_{qq' \to h} / \sigma^{SM}_{gg \to h}) \sim 8$\% and therefore do not qualitatively affect the conclusions.

\subsection{Inclusive di-photon}

The inclusive di-photon rate, on its own, is not necessarily a powerful discriminant of new physics. Nonetheless, as we will see below, it may prove useful in conjunction with precise measurements of certain exclusive channels. 

\begin{figure}[h!!] 
   \centering
   \includegraphics[width=3in]{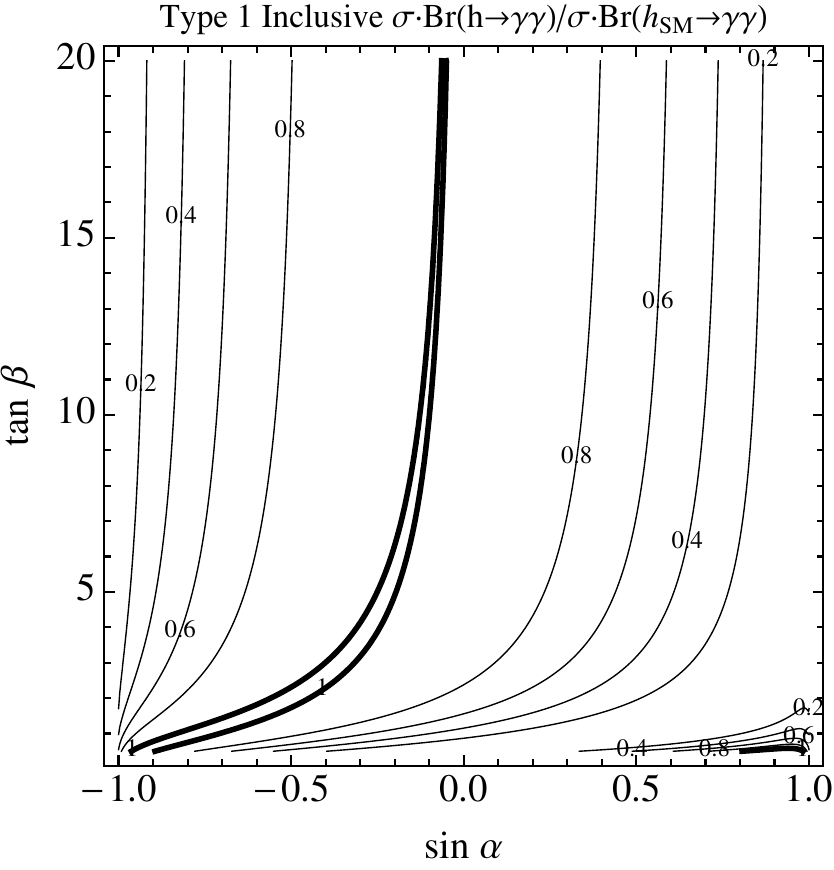} 
    \includegraphics[width=3in]{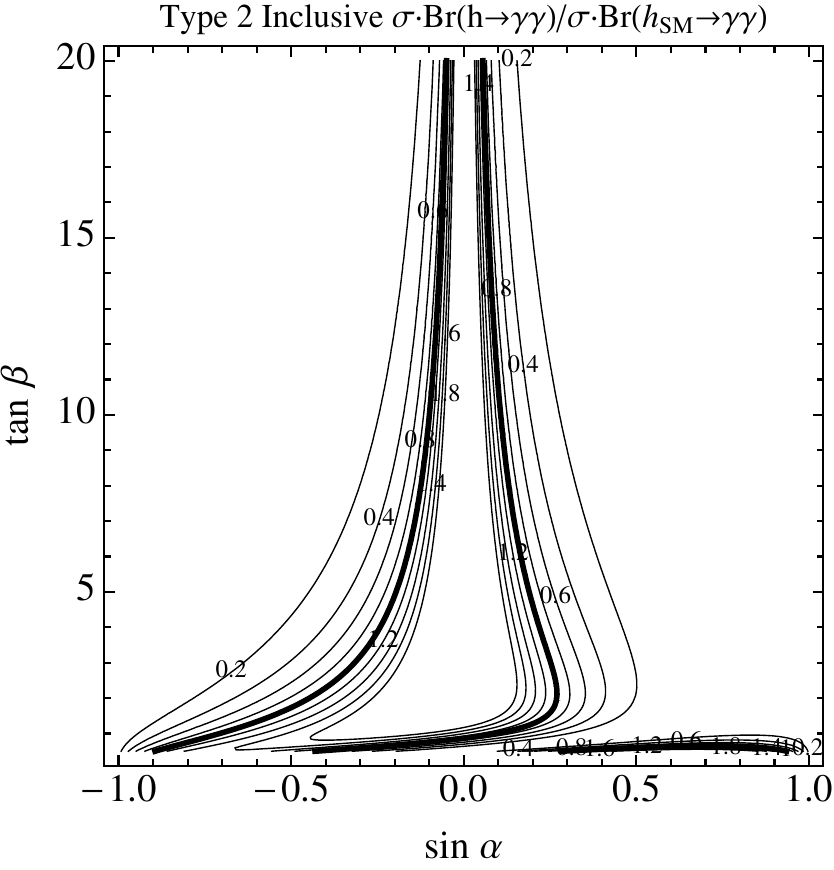} 
   \caption{Contours of the inclusive 
   $\sigmabr (h \to \gamma \gamma) / \sigmabr (h_{SM} \to \gamma \gamma )$  
   for the SM-like Higgs boson with $m_h = 125$ GeV 
   as a function of $\sin \alpha$ and $\tan \beta$ in Type 1 2HDM (left) and Type 2 2HDM (right). 
   The Type 3 model is parametrically similar to Type 1, while Type 4 is similar to Type 2. 
   Thick black lines denote the SM value.  }
   \label{fig:inclusivegammagamma}
\end{figure}

We show contours of the inclusive $\sigma \! \cdot \! {\rm Br}$ for $h \to \gamma \gamma$ in Type 1 and Type 2 2HDM relative to the Standard Model $\sigma \! \cdot \! {\rm Br}$ 
in Fig.~\ref{fig:inclusivegammagamma}; the contours for Type 3 and Type 4 2HDM are respectively similar. In Type 1 and Type 3 2HDM, the contours of the inclusive di-photon $\sigma \! \cdot \! {\rm Br}$ largely track the $g_{hVV} = \sin(\beta - \alpha)$ coupling, rather than the total width. 
Although the branching ratio ${\rm Br}(h \to \gamma \gamma)$ 
increases in inverse proportionality to the total width, the $gg \to h$ 
production rate is directly proportional 
(in its functional dependence on the angles $\alpha, \beta$) 
to the dominant $h \to b \bar b$ partial width, since the $ggh$ coupling is proportional to the $h t \bar{t}$ coupling and both $g_{ht\bar{t}}, g_{h b \bar{b}} \propto \cos \alpha / \sin \beta$. 
As such, an increase in the di-photon branching ratio due to shrinking width is largely offset by a decrease in the production rate. 
Thus the dominant effect on the di-photon inclusive 
$\sigma \! \cdot \! {\rm Br}$ comes from changes in the $hVV$ coupling, which directly affects the partial width $\Gamma(h \to \gamma \gamma)$ since the coupling of the 
Higgs to photons is dominated by a $W$ boson loop. 
Since the $hVV$ coupling saturates at the Standard Model value, 
this suggests that the inclusive di-photon $\sigma \! \cdot \! {\rm Br}$ in Type 1 and Type 3 models is typically bounded from above by the Standard Model rate.  In contrast, in Type 2 and Type 4 2HDM, the contours of the inclusive 
di-photon $\sigma \! \cdot \! {\rm Br}$ largely track the inverse total width. In these theories the $gg \to h$ production rate and total width are not strongly correlated, so that ${\rm Br}(h \to \gamma \gamma)$ may increase as the total width drops, while the production rate $\sigma(gg \to h)$ remains fixed. Since the total width may grow arbitrarily small as $\Gamma(h\to b \bar b)$ decreases, 
the inclusive di-photon $\sigma \! \cdot \! {\rm Br}$ in Type 2 and Type 4 2HDM may be many times the Standard Model rate.

\subsection{Inclusive $VV^*$}

\begin{figure}[h] 
   \centering
   \includegraphics[width=3in]{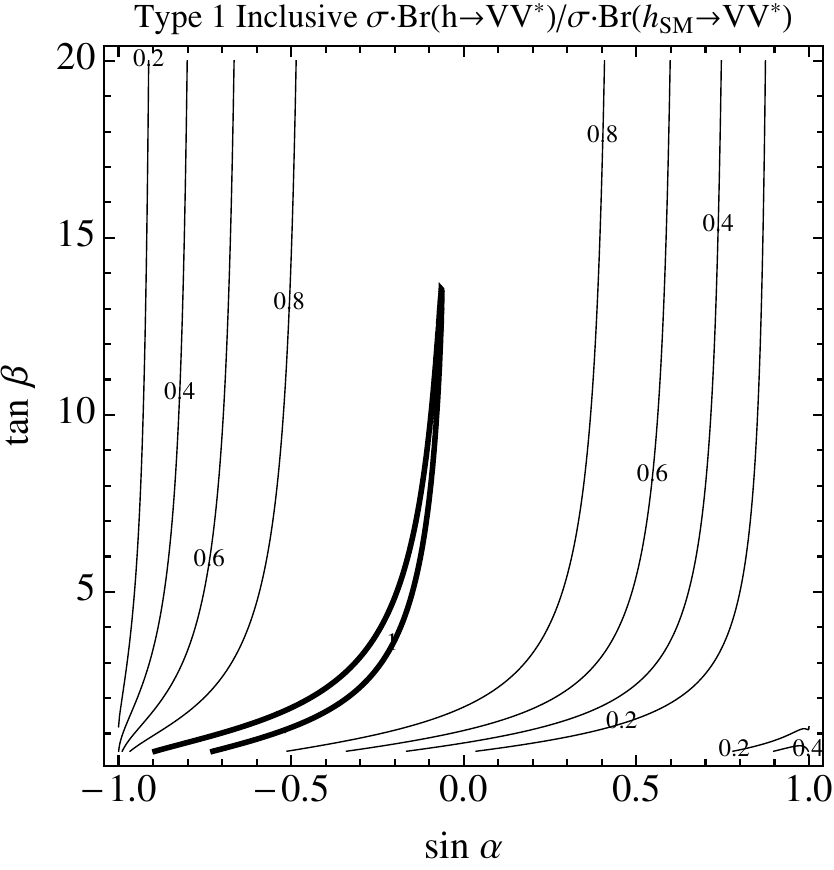} 
     \includegraphics[width=3in]{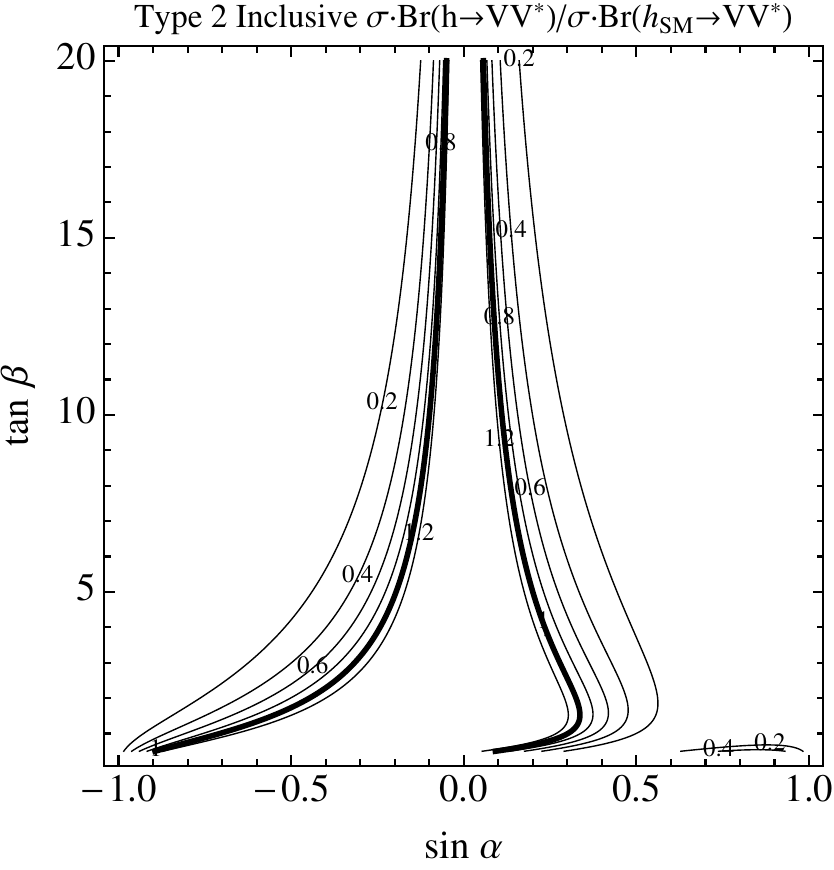} 

  \caption{
  Contours of the inclusive 
   $\sigmabrr (h \to VV^*) / \sigmabrr (h_{SM} \to VV^*)$  
   for the SM-like Higgs boson with $m_h=125$ GeV
   as a function of $\sin \alpha$ and $\tan \beta$ in Type 1 2HDM (left) and Type 2 2HDM (right). 
   The Type 3 model is parametrically similar to Type 1, while Type 4 is similar to Type 2. 
   Thick black lines denote the SM value. 
  }
   \label{fig:inclusiveVV}
\end{figure}

The parametric scaling of the inclusive $h \to VV^*$ $\sigma  \cdot   {\rm Br}$ is quite similar to that of the inclusive di-photon rate, as shown in Fig.~\ref{fig:inclusiveVV}. In the case of Type 1 and Type 3 models, 
the inclusive $\sigma \! \cdot \! {\rm Br}$ again decouples from the total width and largely tracks the partial width $\Gamma(h \to VV^*)$. Whereas the partial width in case of the inclusive di-photon 
$\sigma \! \cdot \!  {\rm Br}$ is a function of both $hVV$ and $ht \bar t$ tree-level couplings, here it is simply a function of the $hVV$ coupling, and so the dependence is somewhat sharper as a function of $\alpha$ and $\beta$. In Type 2 and Type 4 2HDM, as before, the inclusive $\sigma \! \cdot \! {\rm Br}$ tracks the inverse width.

\subsection{Inclusive ratios}

\begin{figure}[h] 
   \centering
   \includegraphics[width=3in]{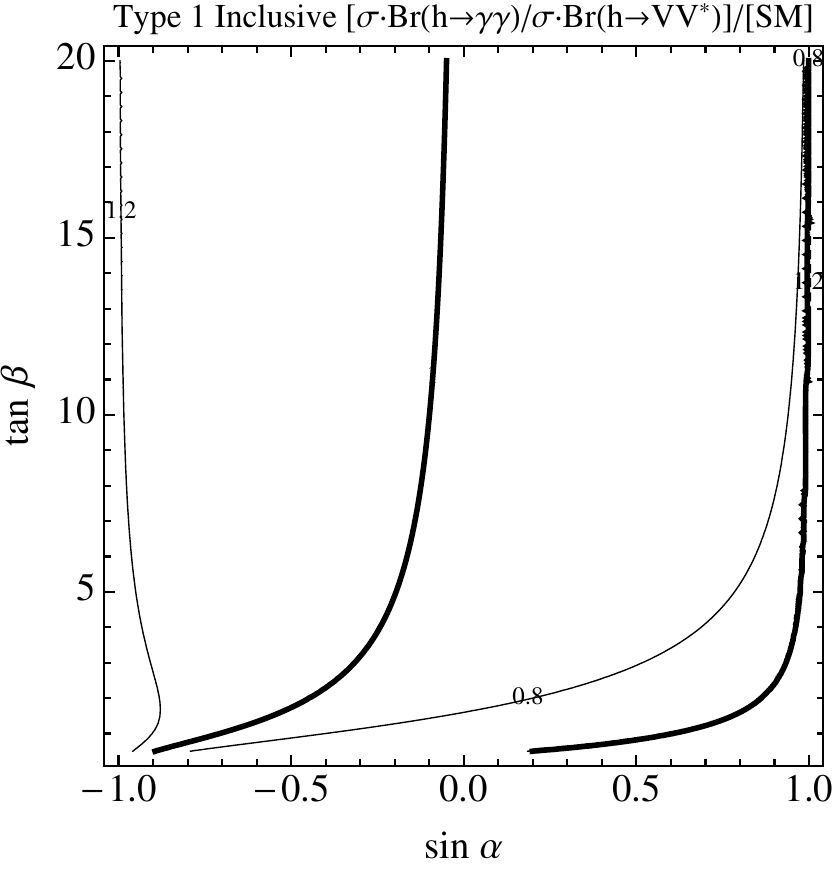} 
     \includegraphics[width=3in]{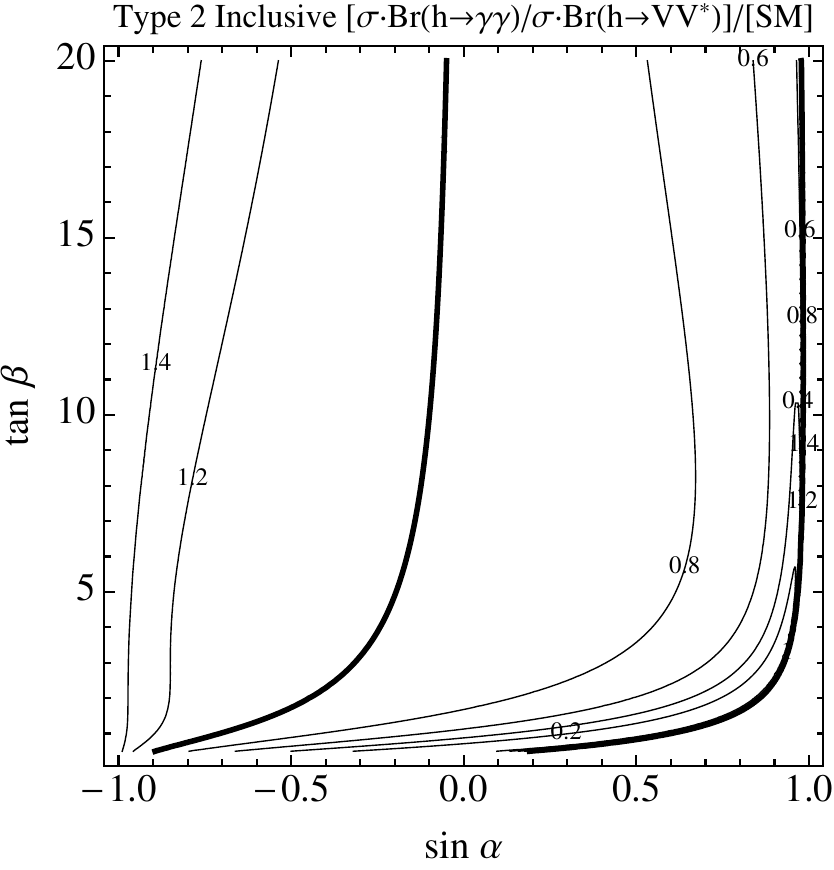} 
     
  \caption{
   Contours of the inclusive ratio 
   $[\sigmabr (h \to \gamma \gamma) / \sigmabr (h \to VV^*)]/
      [\sigmabr (h_{SM} \to \gamma \gamma) / \sigmabr (h_{SM} \to VV^*)]$  
   for the SM-like Higgs boson with $m_h = 125$ GeV
   as a function of $\sin \alpha$ and $\tan \beta$ in Type 1 2HDM (left) and Type 2 2HDM (right). 
   The Type 3 model is parametrically similar to Type 1, while Type 4 is similar to Type 2. 
   Thick black lines denote the SM value. 
 }
   \label{fig:ggVVratio}
\end{figure}

In addition to the inclusive $\sigma \! \cdot \! {\rm Br}$ themselves, we may consider measurements of the inclusive $\gamma \gamma / VV^*$ ratio, shown in Fig.~\ref{fig:ggVVratio}. Dependence on the $ggh$ coupling largely drops out of this ratio, as do many systematics. Given the considerable statistics available in the inclusive channels, this is likely to be the first ratio observed with any meaningful accuracy at the LHC. However, in a 2HDM this inclusive ratio is not particularly sensitive to deviations from SM couplings, due to the fact that both the $h \gamma \gamma$ and $h VV^*$ partial widths scale in large part with the $hVV$ coupling. Although the $h \gamma \gamma$ coupling obtains contributions from both $W$ and top loops, the $W$ contribution dominates; for an SM Higgs at 125 GeV the contribution from the top loop only amounts to $\sim 28\%$ of the total $h \gamma \gamma$ loop amplitude. As such, the variation in this inclusive ratio as a function of $\alpha, \beta$ is quite mild.

\pagebreak

\section{SM-like Higgs exclusive production}
\label{exclusive}

Although the cross section for exclusive channels beyond gluon fusion is more than an order of magnitude smaller than the inclusive cross section, the relatively lower backgrounds mean that many exclusive channels should eventually be observed. The most compelling exclusive channels are VBF and $Vh$ associated production, with Standard Model cross sections at 7-8 TeV ranging from several hundred femtobarns in the case of $Zh$ to more than one picobarn in the case of VBF.
And while the cross section for $t \bar{t} h$ associated production is considerably smaller, 
it may also be observable in certain channels. 

Exclusive channels are compelling not only because of the relatively low backgrounds and isolated production couplings, but also because they admit searches for Higgs final states that are inaccessible in inclusive measurements. In this respect, vector boson fusion with $h \to \tau \tau$ is particularly useful, both for differentiating 2HDM from the Standard Model Higgs and for distinguishing various types of 2HDM from each other. Note that the parametric behavior of several exclusive 
$\sigma \! \cdot \! {\rm Br}$ are identical in 2HDM extensions of the Higgs sector; both vector boson fusion and $Vh$ associated production arise from the same tree-level $hVV$ coupling, so these exclusive production modes scale identically and may be treated collectively. Likewise, for exclusive channels there is less ambiguity relating to adding channels with unknown acceptances.

\subsection{Exclusive VBF/$Wh$/$Zh$ di-photon}

The exclusive VBF di-photon $\sigma \! \cdot \! {\rm Br}$ is a primary 
discriminant between 2HDM and the Standard Model Higgs. 
The inclusive and VBF di-photon signals are observable 
at low integrated luminosity, and a comparison of inclusive and exclusive $\sigma \! \cdot \! {\rm Br}$ 
may provide one of the earliest indications of discrepant Higgs couplings. 

\begin{figure}[htbp] 
   \centering
   \includegraphics[width=3in]{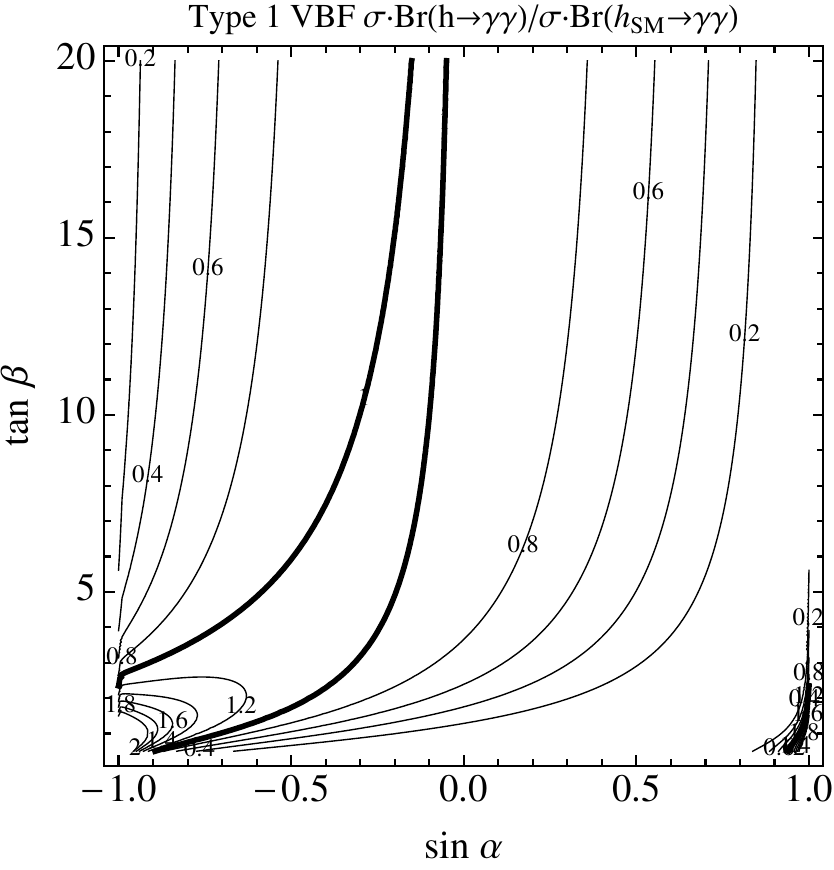} 
      \includegraphics[width=3in]{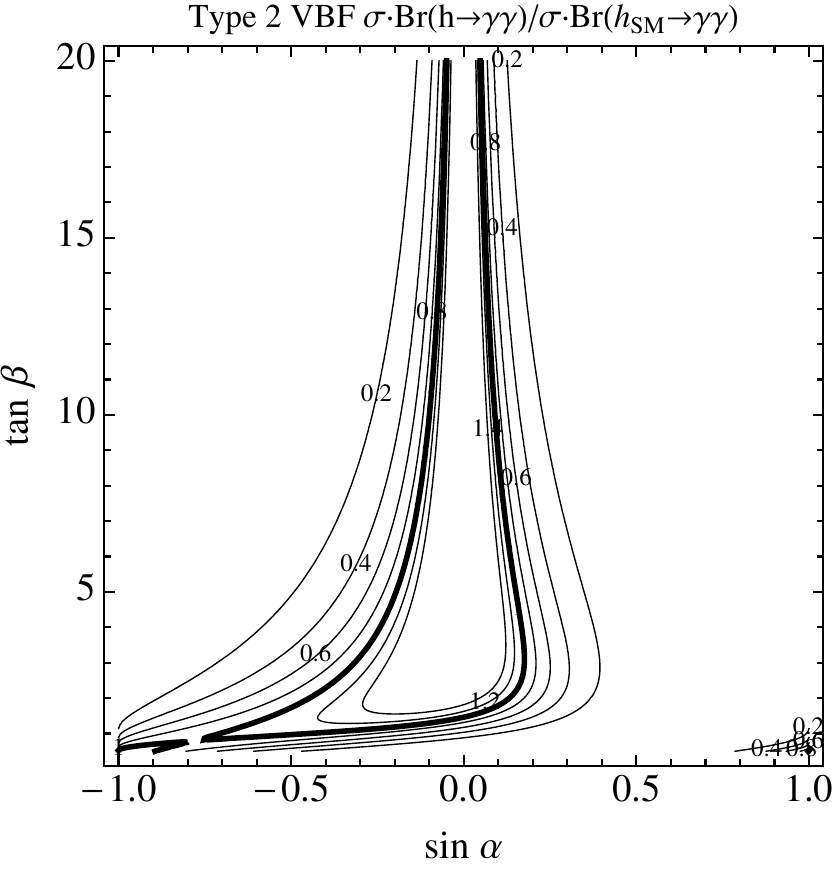} 
   \caption{
   Contours of 
   $\sigmabrr ({\rm VBF}~{\rm or}~Vh \to \gamma \gamma ) / \sigmabrr ({\rm VBF}~{\rm or}~Vh_{SM} \to \gamma \gamma )$  
   for the SM-like Higgs boson  
   as a function of $\sin \alpha$ and $\tan \beta$ in Type 1 2HDM (left) and Type 2 2HDM (right). 
   The Type 3 model is parametrically similar to Type 1, while Type 4 is similar to Type 2. 
   Thick black lines denote the SM value. 
  }
   \label{fig:vbfdi-photon}
\end{figure}

\begin{figure}[htbp] 
   \centering
   \includegraphics[width=3in]{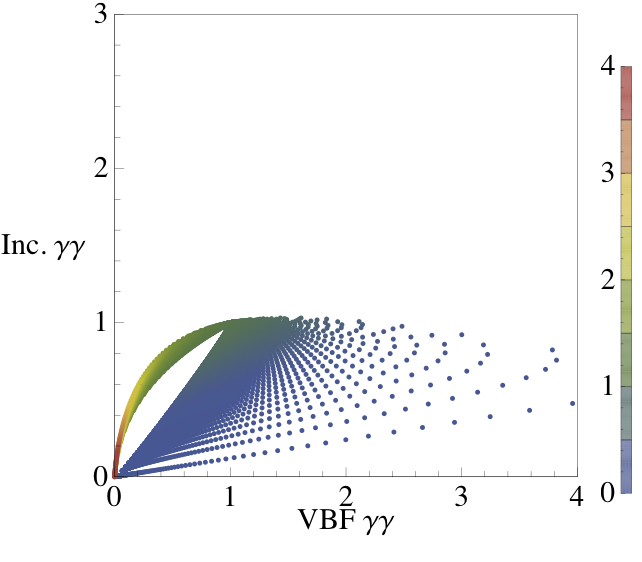} 
      \includegraphics[width=3in]{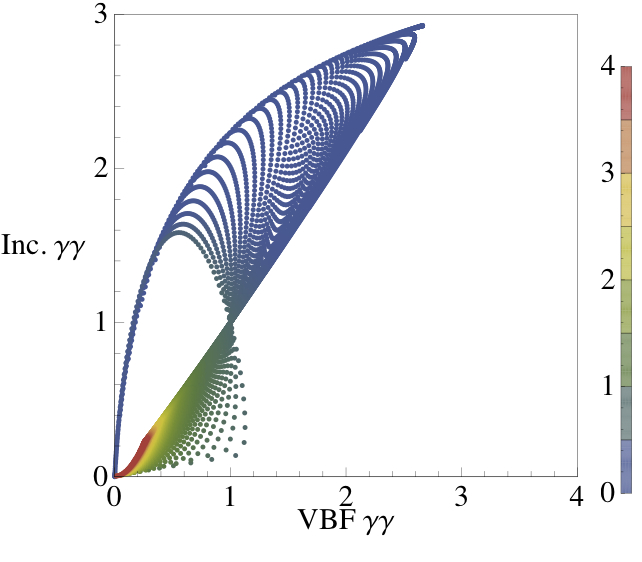} 
   \caption{
   Scatter plot of the inclusive 
   $\sigmabrr (h \to \gamma \gamma ) / \sigmabrr (h_{SM} \to \gamma \gamma )$
   as a function of 
   $\sigmabrr ({\rm VBF}~{\rm or}~Vh \to \gamma \gamma ) / \sigmabrr ({\rm VBF}~{\rm or}~Vh_{SM} \to \gamma \gamma )$ 
       for the SM-like Higgs boson  
       with $m_h = 125$ GeV
   as a function of $\sin \alpha$ and $\tan \beta$ in Type 1 2HDM (left) and Type 2 2HDM (right). 
      The Type 3 model is parametrically similar to Type 1, while Type 4 is similar to Type 2. 
      The points are colorized according to 
      the SM-like Higgs total width relative to the Standard Model Higgs,  
      $\Gamma (h \to {\rm All}) / \Gamma (h_{SM} \to {\rm All} )$, 
            making clear that $\sigmabrr ({\rm VBF}~{\rm or}~Vh \to \gamma \gamma )$
            is anti-correlated with the total width in both 2HDMs.     
            The points are taken from a uniformly spaced grid in $-1 \leq \sin \alpha \leq 0$ and 
            $0.5 \leq \tan \beta \leq 10$ with spacing $\Delta( \sin \alpha) = 0.01$ and 
   $\Delta (\tan \beta) = 0.1$. 
         %
  }
   \label{fig:vbfdi-photonscatter}
\end{figure}

Ratios of the exclusive VBF di-photon $\sigma \! \cdot \! {\rm Br}$ 
relative to the Standard Model are shown in Fig.~\ref{fig:vbfdi-photon}. 
In Type 1 and 3 2HDM, the exclusive di-photon $\sigma \! \cdot \! {\rm Br}$ 
largely tracks the inclusive rate, except in the case of low $\tan \beta$ 
and large mixing, $-0.5 \gtrsim \sin \alpha \gtrsim -1$. 
In this region of parameter space, the VBF di-photon $\sigma \! \cdot \! {\rm Br}$ 
may be significantly enhanced, while the inclusive $\sigma \! \cdot \! {\rm Br}$ is consistent with the SM expectation. The reason for this enhancement is that here the total width 
of the Higgs drops well below the Standard Model value, raising the di-photon branching ratio. 
The $ggh$ coupling drops at the same rate as the width, 
keeping the inclusive $\sigma \! \cdot \! {\rm Br}$ more or less constant, but the 
$hVV$ coupling remains large (since $\sin^2(\beta - \alpha) \sim 1$), so the VBF di-photon $\sigma \! \cdot \! {\rm Br}$ is parametrically enhanced.
 Thus the inclusive $\sigma \! \cdot \! {\rm Br}$ saturates at the SM value, while the VBF di-photon $\sigma \! \cdot \! {\rm Br}$ may be enhanced over the SM $\sigma \! \cdot \! {\rm Br}$ by as much as a factor of four.  In contrast, in  Type 2 and 4 2HDM the inclusive and VBF di-photon $\sigma \! \cdot \! {\rm Br}$ are strongly correlated, and any enhancement or diminuition in the VBF di-photon $\sigma \! \cdot \! {\rm Br}$ 
 points to equally enhanced inclusive rate. Unlike in the previous case, the decrease in width is parametrically distinct from the $ggh$ effective coupling, so both the inclusive and VBF di-photon 
 $\sigma \! \cdot \! {\rm Br}$ are enhanced by a lower total width. Thus a large inclusive 
 di-photon $\sigma \! \cdot \! {\rm Br}$ typically points to a Type 2 or Type 4 2HDM, accompanied by a likely enhancement of the VBF di-photon rate. In contrast, a large VBF di-photon $\sigma \! \cdot \! {\rm Br}$ and SM or smaller inclusive di-photon $\sigma \! \cdot \! {\rm Br}$ would be strongly suggestive of a Type 1 or 3 2HDM. This is exemplified in Fig.~\ref{fig:vbfdi-photonscatter}, which shows the relative range of possible inclusive and VBF di-photon signals in 2HDM.

\pagebreak

\subsection{Exclusive VBF/$Wh$/$Zh$ $VV^*$}

Much as in the case of inclusive processes, the exclusive channels with $h \to VV^*$ exhibit a parametric behavior similar to their $h \to \gamma \gamma$ counterparts, albeit with a slightly smaller SM-like region due to the sensitivity of the tree-level production couplings. In this sense, the exclusive channels with massive gauge boson final states provide a good cross-check of the di-photon $\sigma \! \cdot \! {\rm Br}$. For example, an enhanced VBF di-photon $\sigma \! \cdot \! {\rm Br}$ and SM-like inclusive $\sigma \! \cdot \! {\rm Br}$ should be accompanied by correspondingly enhanced VBF and $Vh$ diboson $\sigma \! \cdot \! {\rm Br}$ with SM-like inclusive cross sections. 



\subsection{Exclusive VBF $\tau \tau$}

\begin{figure}[htbp] 
   \centering
   \includegraphics[width=3in]{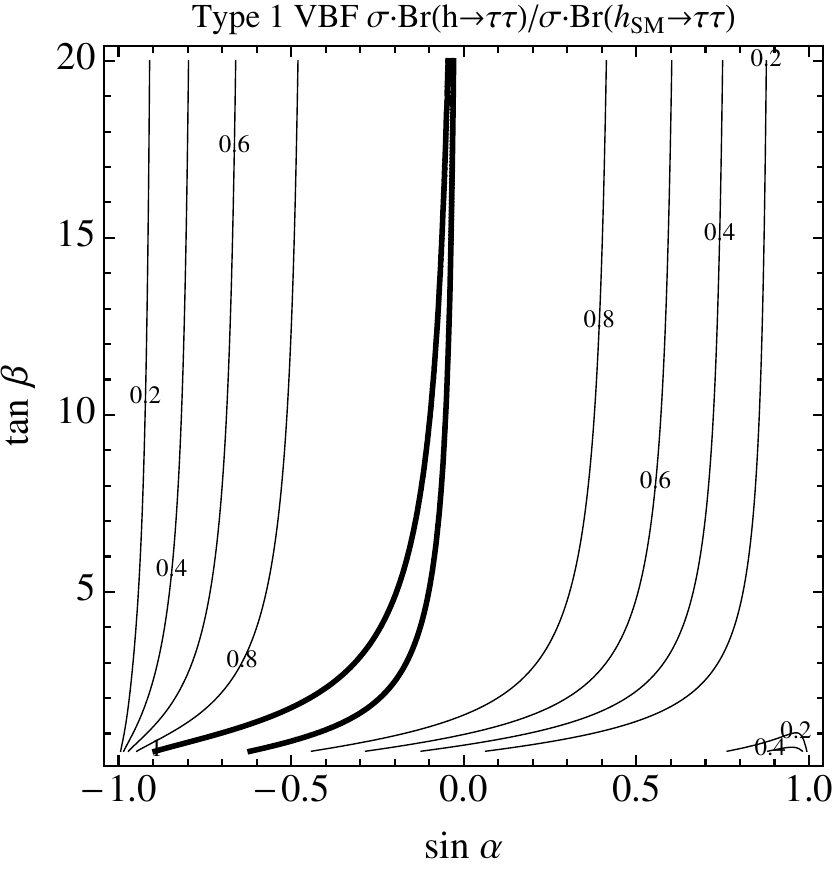} 
      \includegraphics[width=3in]{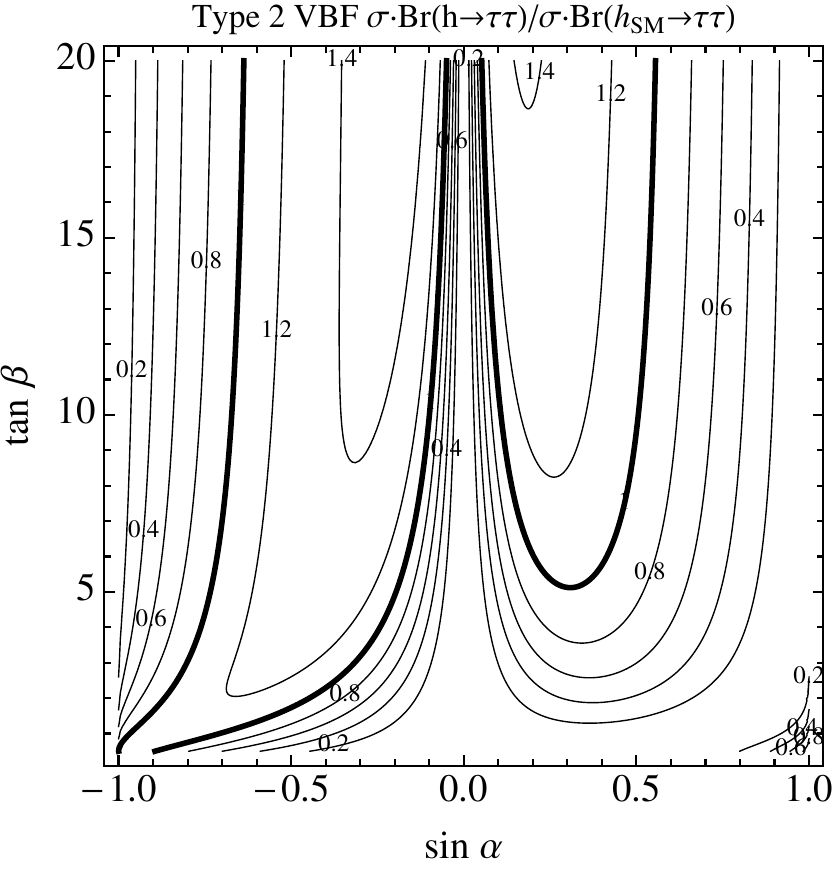} 
      \includegraphics[width=3in]{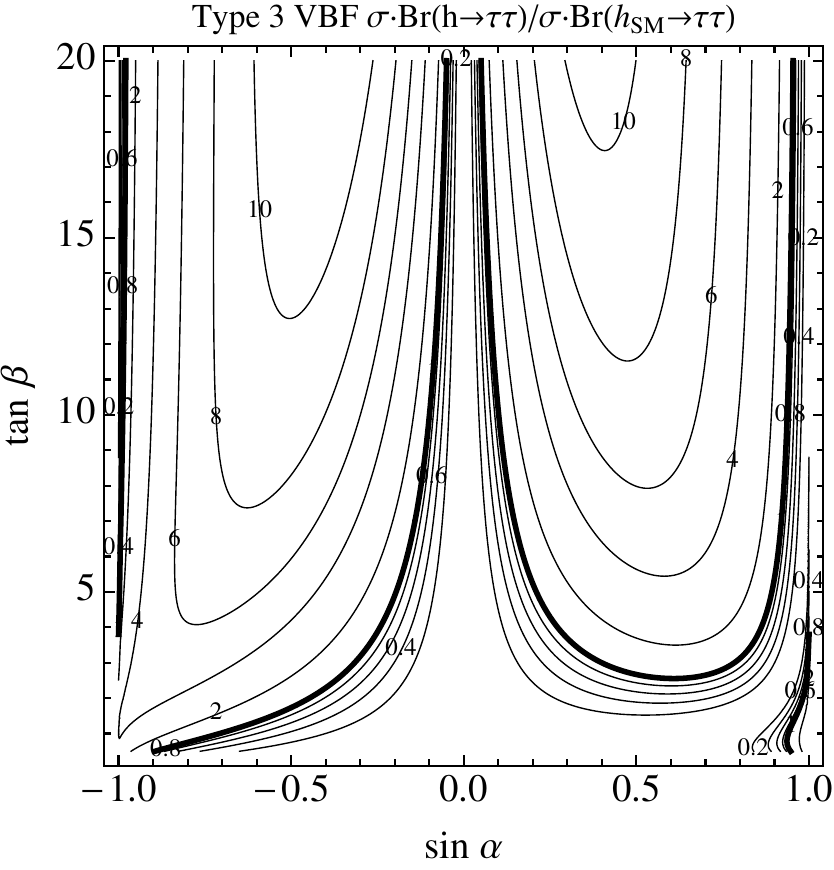} 
      \includegraphics[width=3in]{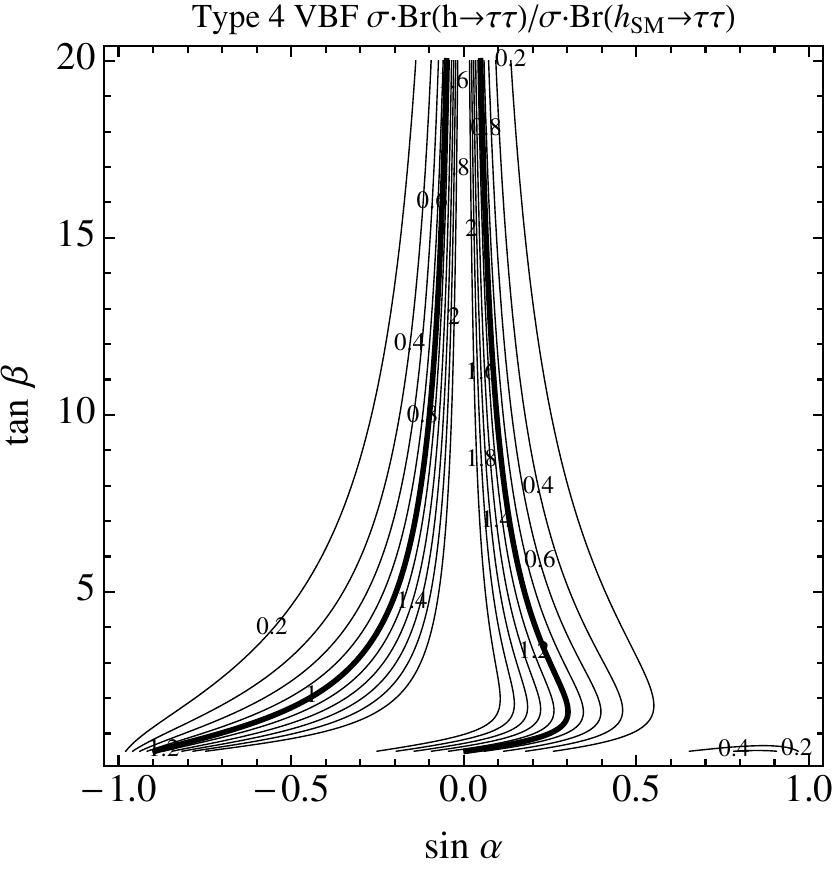} 
   \caption{
     Contours of 
   $\sigmabr ({\rm VBF}~{\rm or}~Vh \to \tau \tau ) / \sigmabr ({\rm VBF}~{\rm or}~Vh_{SM} \to \tau \tau )$  
   for the SM-like Higgs boson  
   as a function of $\sin \alpha$ and $\tan \beta$ in 
   Type 1 2HDM (upper left), Type 2 2HDM (upper right), Type 3 2HDM (lower left), Type 4 2HDM (lower right).   
   Thick black lines denote the SM value. 
   }
   \label{fig:VBFhtautau}
\end{figure}

The VBF $h \to \tau \tau$ $\sigma \! \cdot \! {\rm Br}$ may provide one of the most sensitive discriminants for extended Higgs sectors, both relative to the Standard Model Higgs and among the various types of 2HDM. Ratios of the VBF $h \to \tau \tau$ $\sigma \! \cdot \! {\rm Br}$ relative to the Standard Model expectation are shown in Fig.~\ref{fig:VBFhtautau}; let us briefly consider the different cases. In a Type 1 model, the $h \to \tau \tau$ exclusive production $\sigma \! \cdot \! {\rm Br}$ in VBF is not expected to exceed the Standard Model rate, and in many cases will be significantly smaller. Moreover, the $h \to \tau \tau$ $\sigma \! \cdot \! {\rm Br}$ will typically be SM-like if inclusive di-photon is SM-like. This occurs because the partial width $\Gamma(h \to \tau \tau)$ scales with the total width, so that the branching ratio is largely constant until the width grows small and the $\Gamma(h \to VV)$ partial widths take over. The production cross section lowers as the $hVV$ coupling drops.

In a Type 2 model, the $h \to \tau \tau$ exclusive production $\sigma \! \cdot \! {\rm Br}$ is not expected to exceed SM by more than $\sim 40$ \% for $\tan \beta < 20$, as apparent in Fig.~\ref{fig:VBFhtautau}. If the inclusive di-photon $\sigma \! \cdot \! {\rm Br}$ is SM-like or larger, the $h \to \tau \tau$ exclusive $\sigma \! \cdot \! {\rm Br}$ will be SM-like or smaller. This occurs because the $h \tau \tau$ coupling is parametrically identical to the $h b \bar b$ coupling, which dominates the total width; if the total width is suppressed, leading to enhanced inclusive di-photon signals, the $h \to \tau \tau$ branching ratio drops.

In a Type 3 model, the $h \to \tau \tau$ exclusive  may be many times the SM value, independent of the inclusive di-photon rate. As before, the inclusive di-photon $\sigma \! \cdot \! {\rm Br}$ saturates at the SM value due to the countervailing effects of the width and the $ggh$ coupling. However, in contrast to the Type 1 model, in the Type 3 2HDM the $\Gamma(h \to \tau \tau)$ partial width grows as the total width drops, so that the branching ratio may be considerably enhanced. 

Finally, in a Type 4 model, the $h \to \tau \tau$ $\sigma \! \cdot \! {\rm Br}$ tracks the inclusive di-photon $\sigma \! \cdot \! {\rm Br}$ closely. In Fig.~\ref{fig:VBFhtautau} we see it may greatly exceed the SM value, but an enhanced $\sigma \! \cdot \! {\rm Br}$ is typically accompanied by an equally enhanced inclusive di-photon rate. Again, this arises because the $\Gamma(h \to \tau \tau)$ partial width grows as the total width drops, enhancing the branching ratio by the same effect that enhances the inclusive di-photon rate.

\begin{figure}[htbp] 
   \centering
   \includegraphics[width=3in]{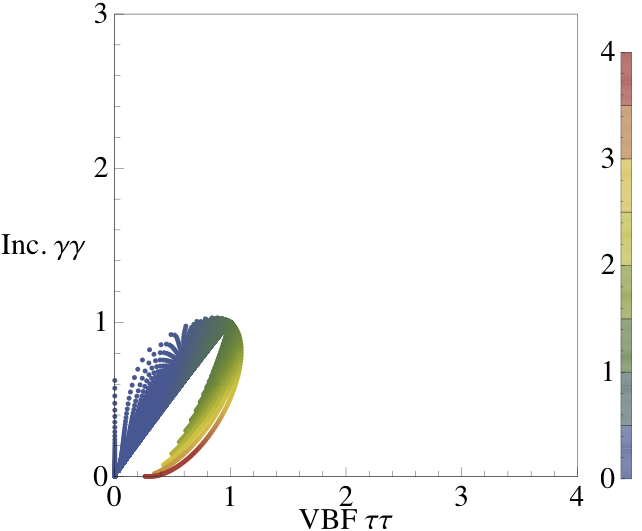} 
      \includegraphics[width=3in]{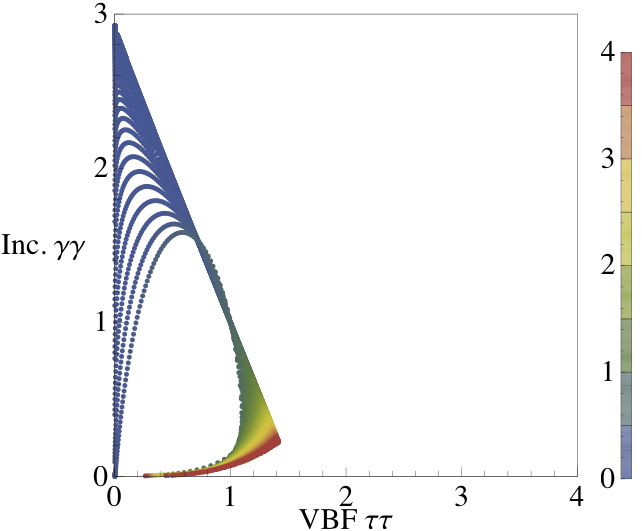} 
          \includegraphics[width=3in]{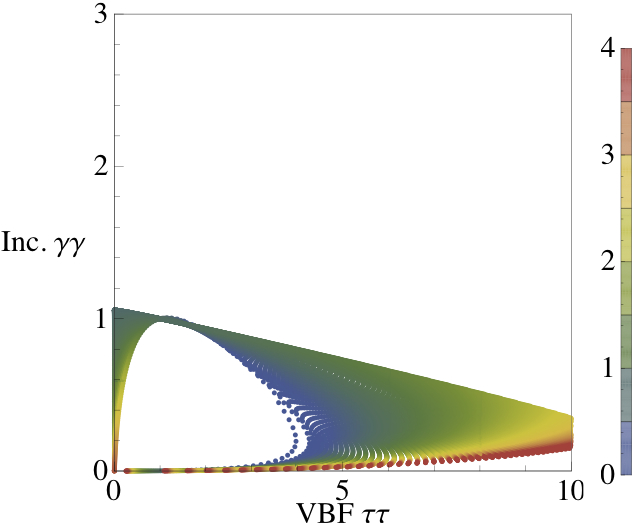} 
              \includegraphics[width=3in]{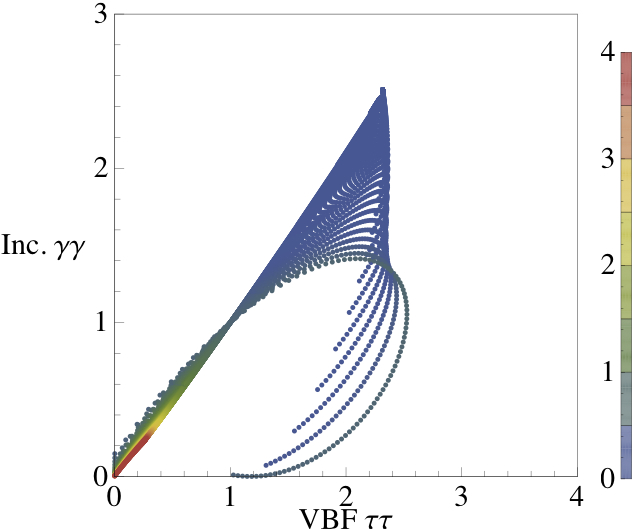} 
   \caption{ 
      Scatter plot of the inclusive $\sigmabr (h \to \gamma \gamma ) / \sigmabr (h_{SM} \to \gamma \gamma )$  as a 
      function of        
      $\sigmabr ({\rm VBF}~{\rm or}~Vh \to \tau \tau ) / \sigmabr ({\rm VBF}~{\rm or}~Vh_{SM} \to \tau \tau )$ 
          for the SM-like Higgs boson  
          with $m_h = 125$ GeV 
   as a function of $\sin \alpha$ and $\tan \beta$ in
   Type 1 2HDM (upper left), Type 2 2HDM (upper right), Type 3 2HDM (lower left), Type 4 2HDM (lower right).   
        The points are colorized according to 
      the SM-like Higgs total width relative to the Standard Model Higgs,  
      $\Gamma (h \to {\rm All}) / \Gamma (h_{SM} \to {\rm All} )$.
          Note that the range for 
           $\sigmabr ({\rm VBF}~{\rm or}~Vh \to \tau \tau ) / \sigmabr ({\rm VBF}~{\rm or}~Vh_{SM} \to \tau \tau )$ 
           differs for the Type 3 2HDM.             
          The points are taken from a uniformly spaced grid in $-1 \leq \sin \alpha \leq 0$ and 
            $0.5 \leq \tan \beta \leq 10$ with spacing $\Delta( \sin \alpha) = 0.01$ and 
   $\Delta (\tan \beta) = 0.1$. 
 }
   \label{fig:VBFhtautauscatter}
\end{figure}

Now one may begin to see clearly the way in which patterns of exclusive $\sigma \! \cdot \! {\rm Br}$ may begin to decisively differentiate between various types of 2HDM. As seen in  Fig.~\ref{fig:VBFhtautauscatter}, discrepancies in the inclusive and exclusive di-photon $\sigma \! \cdot \! {\rm Br}$ provide a clear distinction between Type 1,3 and Type 2,4 models. In the former, the inclusive $\sigma \! \cdot \! {\rm Br}$ is Standard Model or smaller while the exclusive di-photon $\sigma \! \cdot \! {\rm Br}$ may be significantly enhanced; in the latter, the inclusive and exclusive $\sigma \! \cdot \! {\rm Br}$ scale together. If these two classes of 2HDM are distinguished by the inclusive and exclusive di-photon $\sigma \! \cdot \! {\rm Br}$, the particular type of 2HDM may be determined by the exclusive ditau rate. For a Type 1 2HDM the ditau $\sigma \! \cdot \! {\rm Br}$ is SM or smaller, while in the Type 3 2HDM it is typically enhanced; the same is true of Type 2 and Type 4 models, respectively. Thus these three measurements -- the inclusive di-photon, exclusive di-photon, and exclusive ditau $\sigma \! \cdot \! {\rm Br}$ -- may conceivably distinguish among 2HDM models in the event of discrepancies from the Standard Model expectations. Fortunately, they will be among the first processes to be observed.

\subsection{Exclusive VBF/$Wh$ $b \bar b$}

It is further useful to consider exclusive channels with $h \to b \bar b$. The parametric scaling in the Type 1 and Type 3 2HDM is analogous to the Type 1 ditau rate, while that of the Type 2 and Type 4 2HDM is analogous to the Type 2 ditau rate. In this sense, the $h \to b \bar b$ and $h \to \tau \tau$ exclusive $\sigma \! \cdot \! {\rm Br}$ provide a further discriminant between the 2HDM types, though the observation of the $h \to b \bar b$ final state is likely to lag the other exclusive channels somewhat.


\pagebreak

\subsection{Exclusive $t\bar t h$ di-photon}

\begin{figure}[htbp] 
   \centering
   \includegraphics[width=3in]{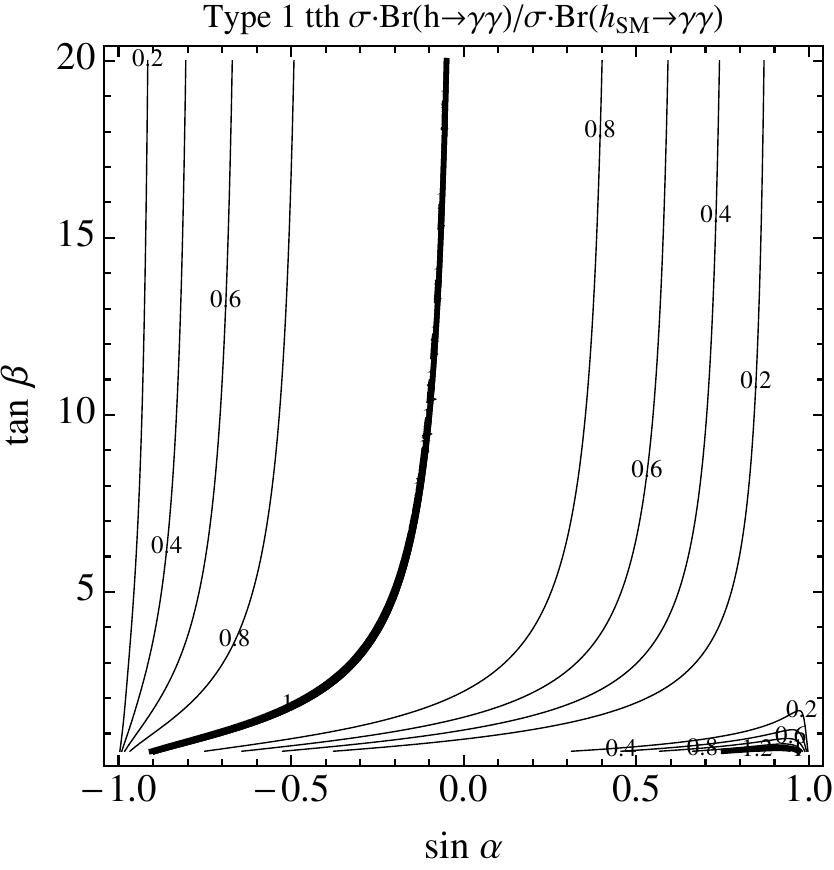} 
      \includegraphics[width=3in]{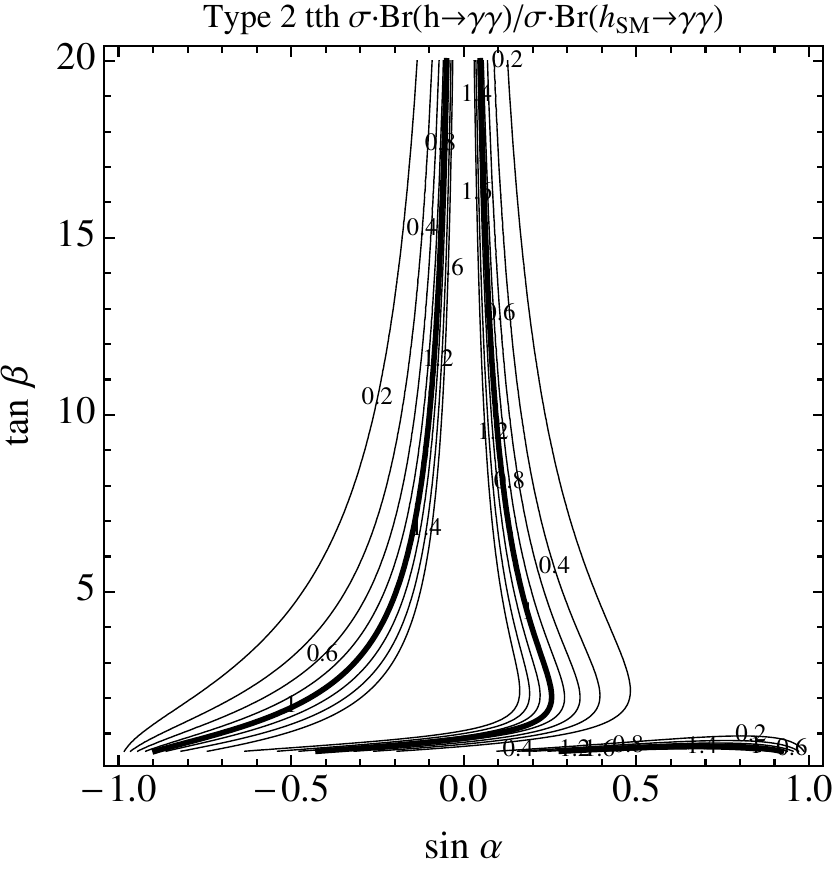} 
   \caption{
   Contours of 
   $\sigmabr (tth \to \gamma \gamma ) / \sigmabr (tth_{SM} \to \gamma \gamma )$  
   for the SM-like Higgs boson  
   as a function of $\sin \alpha$ and $\tan \beta$ in 
   Type 1 2HDM (left) and Type 2 2HDM (right).  
   Thick black lines denote the SM value. 
   }
   \label{fig:tthdi-photon}
\end{figure}

Finally, let us consider the $t\bar th$ di-photon channel, for which contours relative to the Standard Model are shown in Fig.~\ref{fig:tthdi-photon}. The parametric scaling for various 2HDM types is much like the inclusive di-photon scaling, since the production couplings have the same dependence on $\alpha, \beta$ as the $t \bar t$  contribution to the $ggh$ effective coupling. In this sense, $t \bar t h$ provides a good cross-check of the inclusive di-photon rate, though of course the  $t \bar t h$ di-photon signal will be relatively small.  

\pagebreak

\subsection{Exclusive ratios}

Although a side-by-side comparison of various inclusive and exclusive $\sigma \! \cdot \! {\rm Br}$ provides considerable discrimination between various 2HDM types, each channel is accompanied by its own particular systematic errors. These errors reduce the significance of any departures from Standard Model expectations, making it unlikely that decisive observations of new physics can be made with low luminosity for all but the most extreme cases. Thus it is useful to examine certain exclusive ratios. Various systematics drop out of ratios of the same exclusive production channel with different final states, making such ratios sensitive to deviations in Higgs decays.

\begin{figure}[h] 
   \centering
   \includegraphics[width=3in]{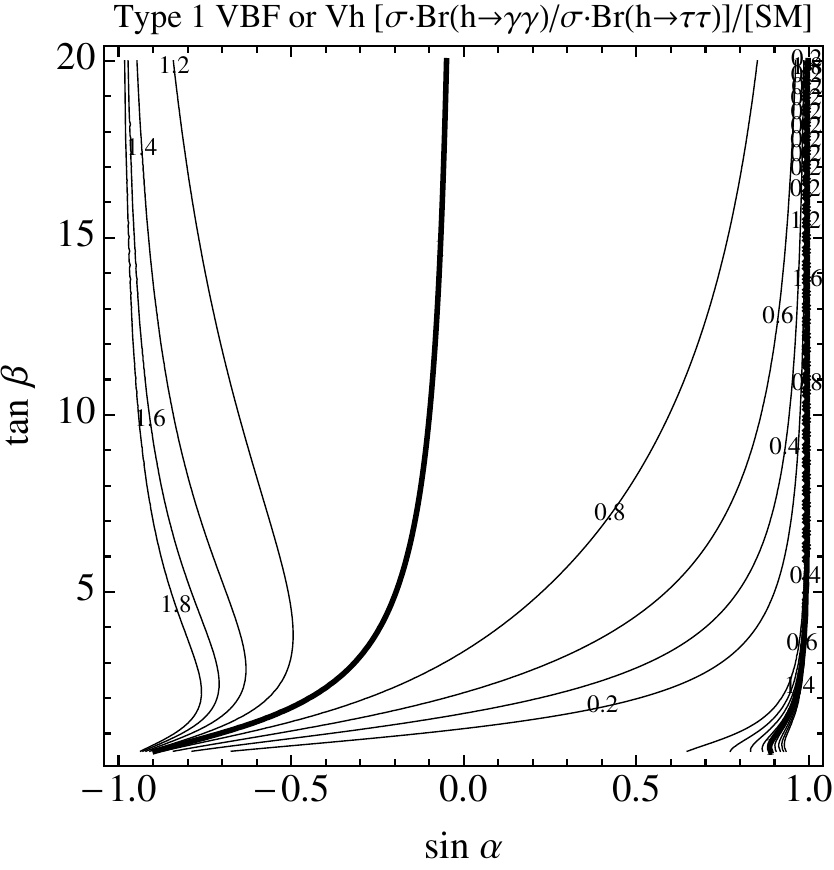} 
     \includegraphics[width=3in]{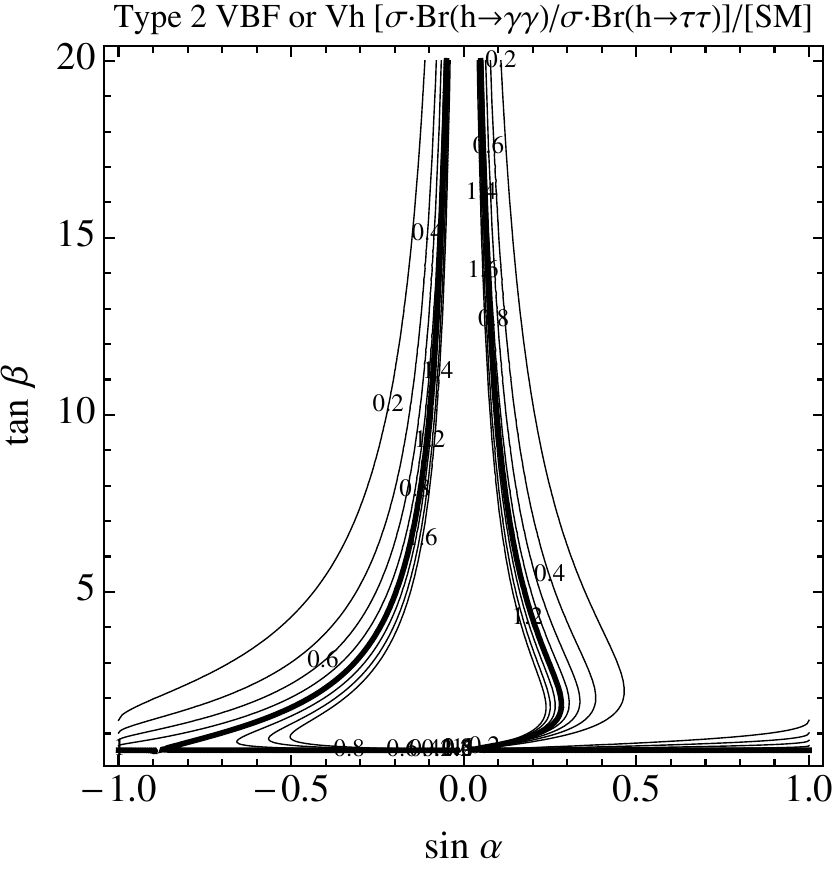} 
     
  \caption{
  Contours of 
   $[\sigmabrr ({\rm VBF}~{\rm or}~Vh \to \gamma \gamma ) / 
      \sigmabrr ({\rm VBF}~{\rm or}~Vh \to \tau \tau )]/
     [\sigmabrr ({\rm VBF}~{\rm or}~Vh_{SM} \to \gamma \gamma ) / 
      \sigmabrr ({\rm VBF}~{\rm or}~Vh_{SM} \to \tau \tau )] $  
   for the SM-like Higgs boson  
   as a function of $\sin \alpha$ and $\tan \beta$ in 
   Type 1 2HDM (left) and  Type 2 2HDM (right).   
   Thick black lines denote the SM value. 
    }
   \label{fig:VBFdi-photonditauratio}
\end{figure}

The parametric scaling of the VBF $\gamma \gamma / VV$ exclusive ratio is much like that of the inclusive $\gamma \gamma / VV$ ratio, and we need not reproduce the figure here. Far more interesting is the VBF $\tau \tau / \gamma \gamma$ exclusive ratio. Unsuprisingly, given the discriminating power of the $h \to \tau \tau$ final state, this exclusive ratio is fairly sensitive, varying rapidly from the Standard Model ratio in all types of 2HDM models as illustrated in Fig.~\ref{fig:VBFdi-photonditauratio}. Finally, the VBF $\tau \tau / VV$ exclusive ratio has similar parametric behavior to the $\tau \tau / \gamma \gamma$ exclusive ratio. This exclusive ratio is particularly attractive, as theory studies of Higgs coupling measurements suggest this will be one of the earliest exclusive ratios to be probed at the LHC \cite{Lafaye:2009vr}.

\pagebreak

\section{Heavy non-SM-like scalars}
\label{heavy}

Thus far we have restricted our attention entirely to the SM-like 
CP even neutral scalar $h$, since its properties initially merit the closest scrutiny. 
However, the other states present in 2HDMs may likewise appear in standard Higgs search 
channels, and their production rates can be 
 constrained by existing Higgs searches. 
Of the various non-SM like Higgs states, 
the CP even neutral scalar $H$ and the CP odd neutral scalar 
$A$ are the most likely to appear in standard Higgs channels. 
The $H$ is produced through both gluon-gluon fusion and 
all of the associated production channels, and could in principle be visible 
in the decay modes $H \to VV$, particularly $H \to WW \to 2 \ell 2 \nu$ and 
$H \to ZZ \to 4 \ell, 2 \ell 2 \nu, 2 \ell 2j, 2 \ell 2 \tau$, 
as well as $H \to \gamma \gamma$. 
With CP conservation, 
the pseudoscalar $A$ has zero tree-level coupling to Standard Model 
gauge bosons, although it does possess the loop-level 
$A \gamma \gamma$ and $Agg$ effective 
couplings. 
As such, the pseudoscalar is produced 
only through gluon-gluon fusion and $t \bar t A$ associated production. 
Likewise, it lacks the $A \to VV$ final states of high-mass standard 
Higgs searches, and so 
among the standard Higgs search channels it could in principle be visible 
in the $A \to \gamma \gamma$ decay mode. 
Note that the $A \to \tau \tau$ decay mode may also be significant, but because it does 
not occur in associated production, it's unlikely to appear in standard channels at low luminosity.

Of course, it is also possible for the non-SM like 
states to manifest themselves primarily in scalar cascade decays, ending in 
Standard Model final states plus the light SM-like Higgs $h$. 
These cascades may appear in standard Higgs channels as an enhancement of associated production processes. For example, the process $gg \to A \to Zh$ may have a sizable cross section, and appears as an enhancement of the $Zh$ exclusive channel (albeit with boosted kinematics). In general, these scalar decays are highly model-dependent; their rates depend on unknown free parameters in the scalar potential 
that determine the Higgs self couplings, 
and on the masses of the various scalar states. 
Insofar as our focus lies on signatures appearing in standard Higgs channels, we will not give these scalar cascade signals further consideration here.  

Needless to say, the production cross sections and branching ratios for the heavy scalars 
depend sensitively on their masses and partial decay widths 
to any non-Standard Model states.  
However, in the limit where their decay widths are dominated by Standard Model final states, 
we can make 
quantitatively accurate statements about the 
cross section times branching ratios of these states relative to the Standard Model 
Higgs of the same mass.  
This is not an unreasonable approximation in many regions of parameter space, and more generally provides an upper limit on the size of signals measurable in standard Higgs channels. 
Once the masses of the heavy Higgses are above the top quark threshold, 
$2m_t$, their branching ratios to Standard Model final states vary slowly since the dominant partial width to $t \bar t$ varies slowly with the scalar mass.

\begin{figure}[h!!] 
   \centering
   \includegraphics[width=3in]{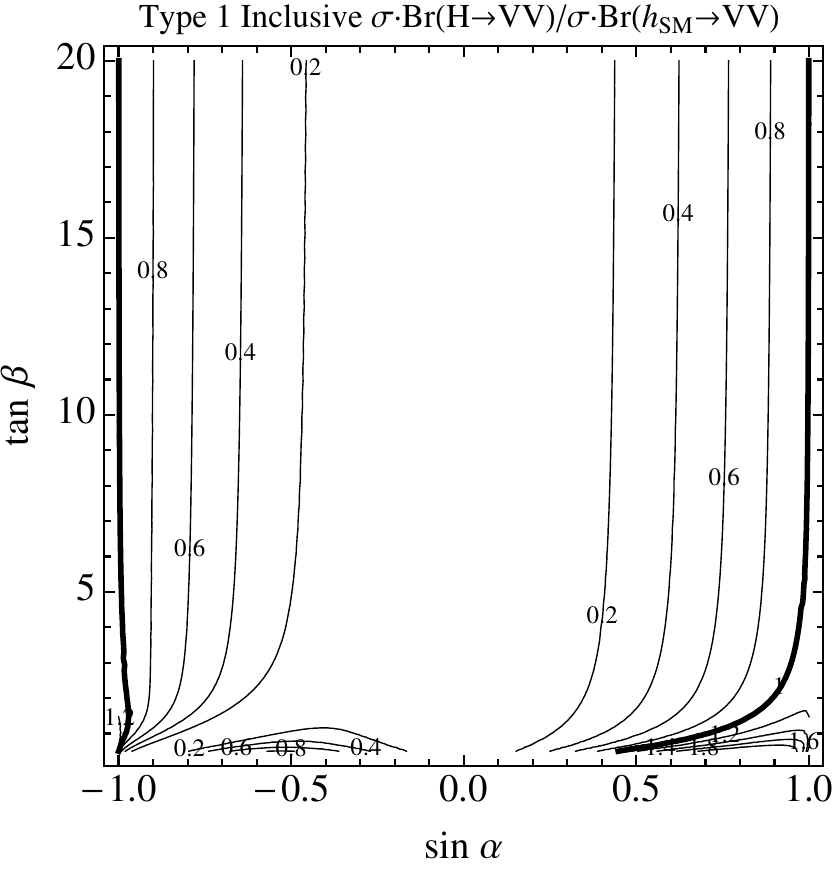} 
   \caption{
   Contours of the inclusive
   $\sigmabr (H \to VV )/ \sigmabr (h_{SM} \to VV )$  
   for the non-SM-like scalar Higgs boson assuming
   decays to Standard Model fermions and gauge bosons only 
   as a function of $\sin \alpha$ and $\tan \beta$ in 
   Type 1 2HDM with $m_H = m_{h_{SM}} = 400$ GeV.
      Thick black lines denote the SM value.
      Contours for other types of 2HDM are similar. 
  }
   \label{fig:Hdi-photon}
\end{figure}

 

\begin{figure}[h!!] 
   \centering
   \includegraphics[width=3in]{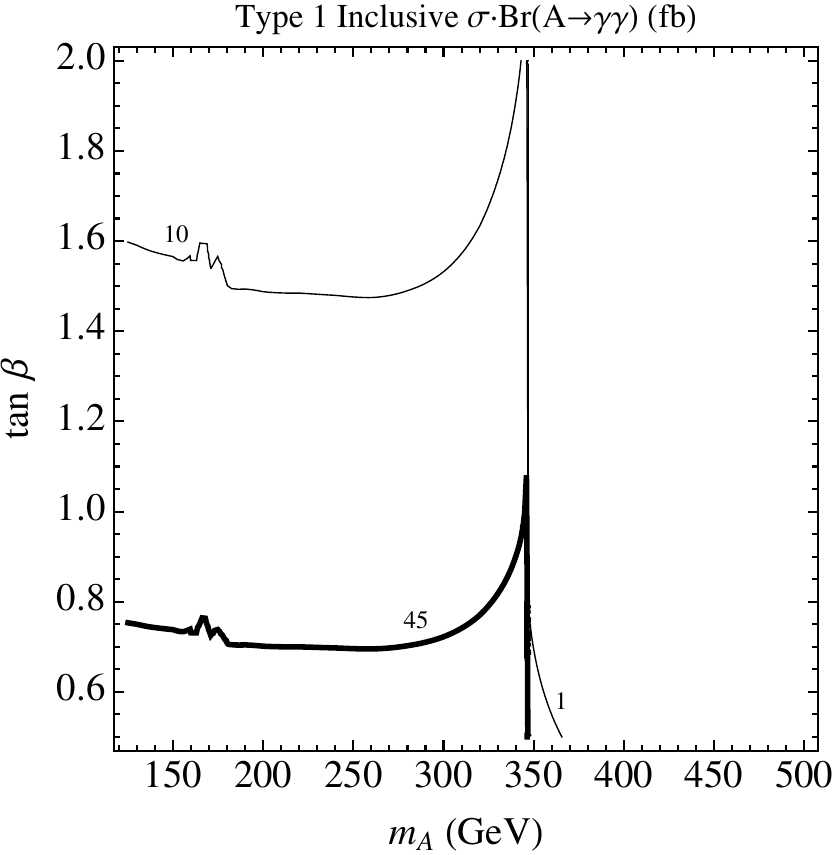} 
      \includegraphics[width=3in]{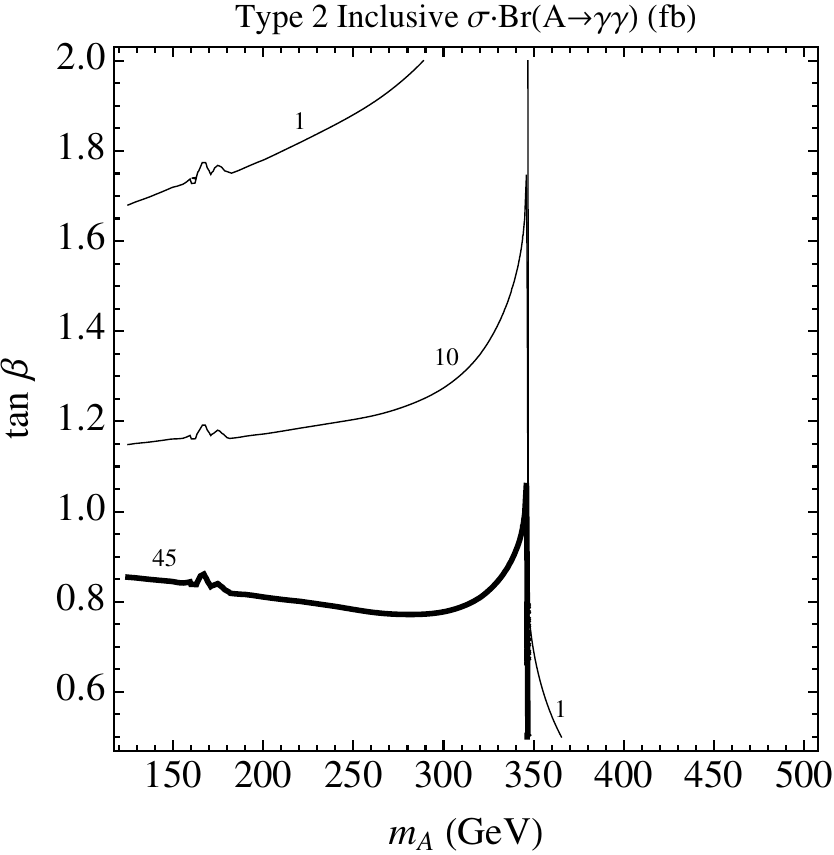} 
 
  \caption{
  Contours of the inclusive
   $\sigmabr (A \to \gamma \gamma ) $ in units of fb
   for 8 TeV $pp$ collisions 
   for the pseudo-scalar Higgs boson  
   as a function of $\tan \beta$ and $m_A$ in 
   Type 1 2HDM (left) and Type 2 2HDM (right)
   assuming
   decays to Standard Model fermions and gauge bosons only.
         For reference the thick black 
         lines are close to the SM value $\sigmabr (h_{SM} \to \gamma \gamma ) \simeq 45$ fb 
         for $m_h = 125$ GeV.     
      Contours for Type 3 2HDM are similar to Type 1, and Type 4 2HDM to Type 2. 
      The rapid drop in $\sigmabr (A \to \gamma \gamma ) $ for $m_A \gsim 2 m_t$ is due to the top quark 
      threshold.  
         %
 }
   \label{fig:Adi-photon}
\end{figure}

The inclusive diboson $\sigma \! \cdot \! {\rm Br}$ for the heavy scalar $H$ 
to $VV$ 
is similar among the various types of 2HDM, as shown in Fig.~\ref{fig:Hdi-photon} for $m_H = 400$ GeV, normalized to the $\sigma \! \cdot \! {\rm Br}$ of a Standard Model Higgs of the same mass. This is due in large part to the fact that the width at high mass is dominated by top and vector boson final states, which have the same parametric scalings in all the types of 2HDM (with a mild exception at large $\tan \beta$ where $h \to b \bar b$ becomes important for the Type 2 and Type 4 2HDM). 
What is particularly apparent in Fig.~\ref{fig:Hdi-photon}
is the SM Higgs-like inclusive $\sigma \! \cdot \! {\rm Br}$ (compared to a SM Higgs at that mass) in the limit $\sin \alpha \to \pm 1$. 
These are exceptionally interesting regions of parameter space, insofar as it is also in this limit that many of the standard channel $\sigma \! \cdot \! {\rm Br}$ of $h$ differ widely from the Standard Model value, particularly at small $\tan \beta$. 
Even away from these limits, the $\sigma \! \cdot \! {\rm Br}$
for $H \to VV$ can be a sizeable fraction of the value for a SM Higgs boson of the same mass. 
So standard Higgs searches for $VV$ final states for masses even well above the 
SM-like 
Higgs mass, can be 
useful probes of 2HDMs.  
The correlated complementary nature of $h \to VV^*$ and $H \to VV$ processes 
in a 2HDM can be see by comparing Figs.~\ref{fig:inclusiveVV} and \ref{fig:Hdi-photon}. 
A sizeable deviation of $\sigma \! \cdot \! {\rm Br}$ for 
$h \to VV^*$ from SM expectations 
implies a sizeable signal for $H \to VV$ 
(relative to a SM Higgs boson of the same mass). 
Conversely good agreement of $\sigma \! \cdot \! {\rm Br}$ for  $h \to VV^*$ 
with the SM expectation implies a reduced signal for $H \to VV$.

\begin{figure}[h!!] 
   \centering
   \includegraphics[width=3in]{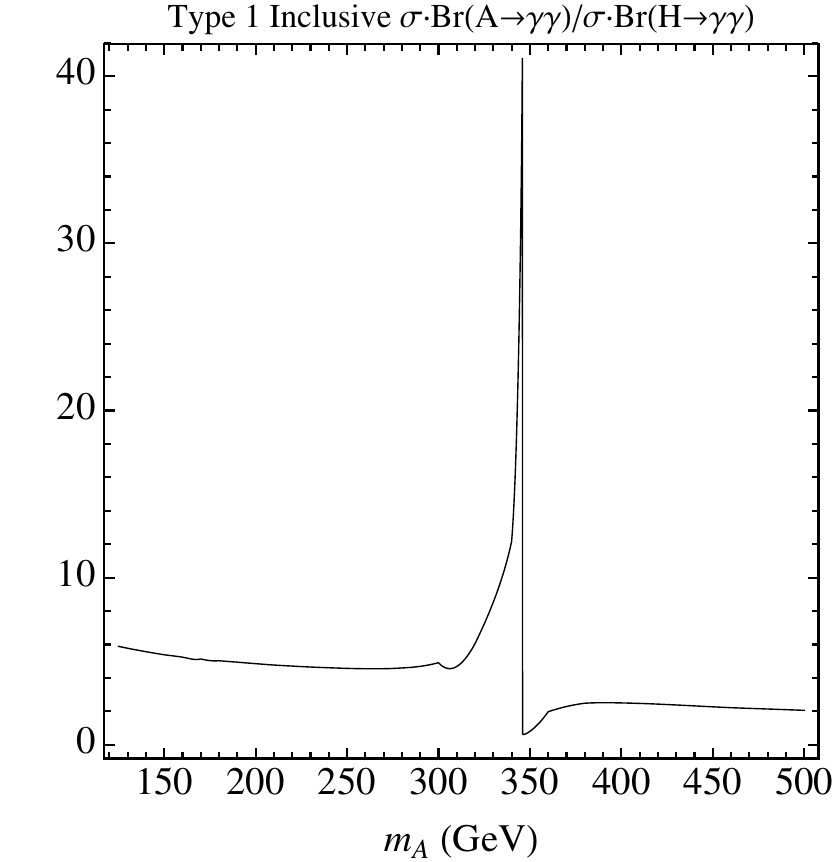} 
      \includegraphics[width=3in]{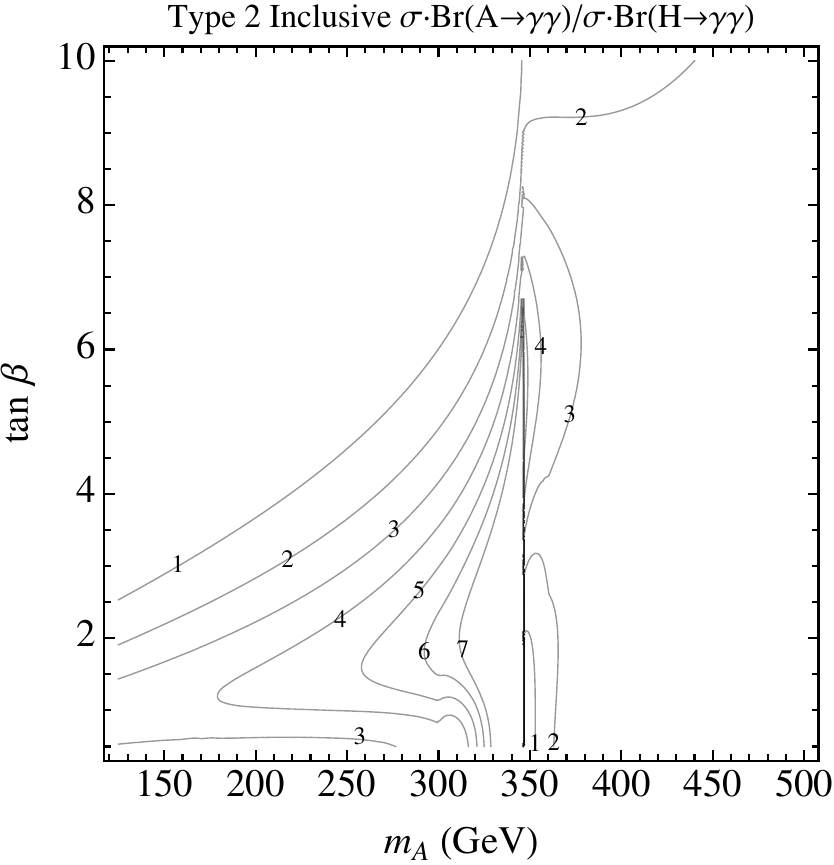} 
 
  \caption{
  The inclusive 
   $\sigmabr (A \to \gamma \gamma ) / \sigmabr (H \to \gamma \gamma )$    
   for the pseudo-scalar and heavy scalar Higgs bosons in the  $\sin^2(\beta - \alpha)=1$ 
   alignment  limit 
   as a function of $m_A=m_H$ in 
   Type 1 2HDM (left) and  
   as a function of $m_A=m_H$ and $\tan \beta$ in
    Type 2 2HDM (right)
   assuming
   decays to Standard Model states only.
   The pseudo-scalar couples only to fermions, and in the alignment 
   limit the heavy scalar also couples only 
   to fermions.
             Contours for Type 3 2HDM are similar to Type 1, and Type 4 2HDM to Type 2. 
      The rapid change in the ratio for $m_A \gsim 2 m_t$ is due to the top quark 
      threshold.  
         %
 }
   \label{fig:AHdi-photon}
\end{figure}

The production and decay modes of the pseudoscalar $A$ depend only on $\tan \beta$ 
through its couplings to fermions,
and are independent of the scalar mixing angle $\alpha$.  
The inclusive 
di-photon $\sigma \! \cdot \! {\rm Br}$ for  $A$
is shown in Fig.~\ref{fig:Adi-photon} as a function of $\tan \beta$ and 
the pseduoscalar mass.  
This  $\sigma \! \cdot \! {\rm Br}$ decreases rapidly with $\tan \beta$, since both the production and decay rates depend on the square of the $t \bar t A$ coupling. 
 But at small $\tan \beta$ the di-photon signals of the pseudoscalar may be several times larger than a Standard Model Higgs of equivalent mass. 
Existing searches for 2HDM focus on $\tau$-lepton 
final states of either the 
charged Higgs $H^\pm$ coming from top quark decay or 
 $A$ and $H$ produced in association with $b$-quarks. 
These searches are sensitive to large $\tan \beta$, and so the inclusive 
$A \to \gamma \gamma$ channel provides an interesting complementary 
probe of low $\tan \beta$.

The di-photon $\sigma \! \cdot \! {\rm Br}$ for the heavy scalar 
$H$ can be similar in magnitude to that of the pseudoscalar $A$, 
though in general it depends on both $\alpha$ and $\beta$. 
However, in the $\sin(\beta - \alpha) = 1$
alignment limit it depends at most on $\tan \beta$ since 
$H$ couples only to fermions in this limit.
And the specific form of the dependence  is rather simple because 
in any of the 2HDMs the magnitude and $\tan \beta$ dependence 
of the coupling of $H$ to a given type of fermion is identical to 
that of the $A$.  
Consider first the ratio  
$\sigma \! \cdot \! {\rm Br}(H \to \gamma \gamma) /  \sigma \! \cdot \! {\rm Br}(A \to \gamma \gamma)$ 
for $H$ and $A$ of the same mass 
in the Type 1 2HDM shown in Fig.~\ref{fig:AHdi-photon}. 
The ratio in the alignment limit in this case is independent of $\tan \beta$ since 
the $H$ and $A$ couplings to all fermions are homogeneous in 
Type 1. 
Below the top threshold 
this ratio depends mainly on the overall coefficient of the fermion 
loop diagrams.  
Above the top threshold the relative di-photon branching 
ratio dips rapidly since the $A \to tt$ decay is $S$-wave 
near threshold, while 
$H \to tt$ is $P$-wave.
In contrast the ratio 
$\sigma \! \cdot \! {\rm Br}(H \to \gamma \gamma) /  \sigma \! \cdot \! {\rm Br}(A \to \gamma \gamma)$ 
in Type 2 2HDM, also shown in Fig.~\ref{fig:AHdi-photon},
does depend on $\tan \beta$ in the alignment limit
because the top and $b$-quark couplings have different $\tan \beta$ 
dependence. 
So in principle both $A$ and $H$ could be observed as two additional di-photon 
resonances beyond that of the SM-like Higgs, 
although the $H$ resonance is generally somewhat weaker than 
the $A$ resonance. 
The search for these two resonances over the full experimentally 
accessible mass range represents an important probe of the 
low $\tan \beta$ region of any of the 2HDMs.  
Observation of these two di-photon resonances would 
give accurate measures of the masses that could in turn yield 
additional information about the Higgs self couplings \cite{Dine:2007xi}.

\section{SM-like Higgs Signal Fits}
\label{fits}

The primary aim of the work presented here is to provide a 2HDM roadmap for potential 
deviations of the signals of a Higgs boson from Standard Model expectations, and the 
correlated  signals 
of additional scalars  
in standard Higgs search channels.  
It is instructive to consider this roadmap  
in the light of current fits to the Standard Model-like Higgs discovery level 
signal observed by ATLAS and CMS. 
For simplicity, we restrict our focus to the CMS results
 for the two most sensitive channels, the inclusive $\sigma \! \cdot \! {\rm Br} (h \to \gamma \gamma) / \sigma \! \cdot \! {\rm Br}(h_{SM} \to \gamma \gamma)$ and $\sigma \! \cdot \! {\rm Br} (h \to ZZ^*) / \sigma \! \cdot \! {\rm Br}(h_{SM} \to ZZ^*)$ \cite{CMS:2012gu}. 

\begin{figure}[h!!] 
   \centering
   \includegraphics[width=3in]{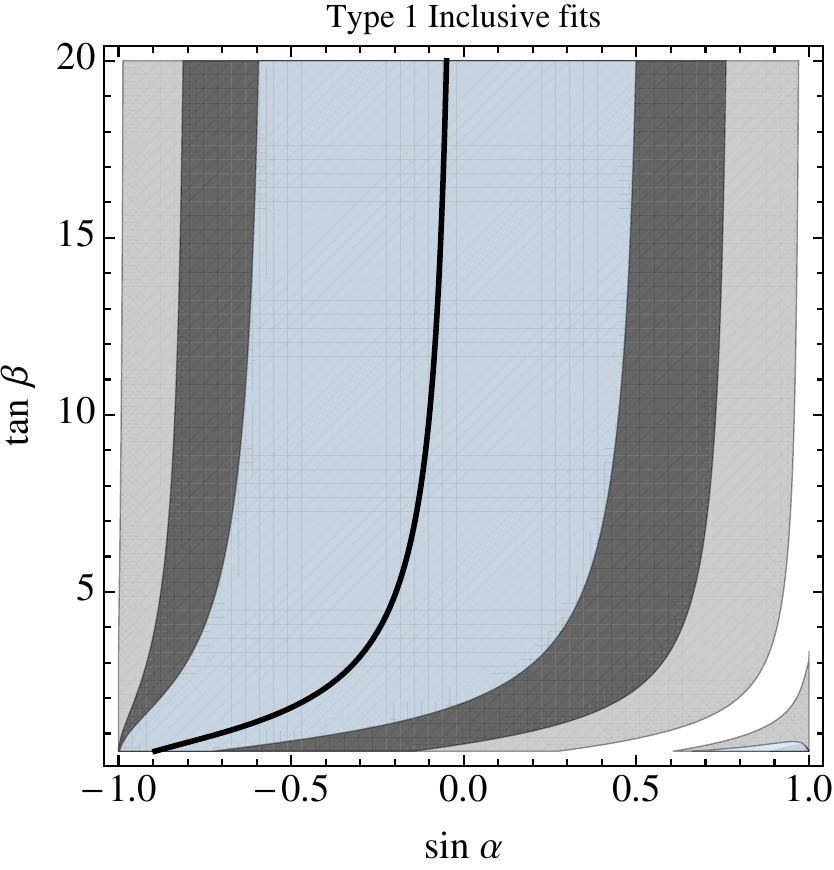} 
      \includegraphics[width=3in]{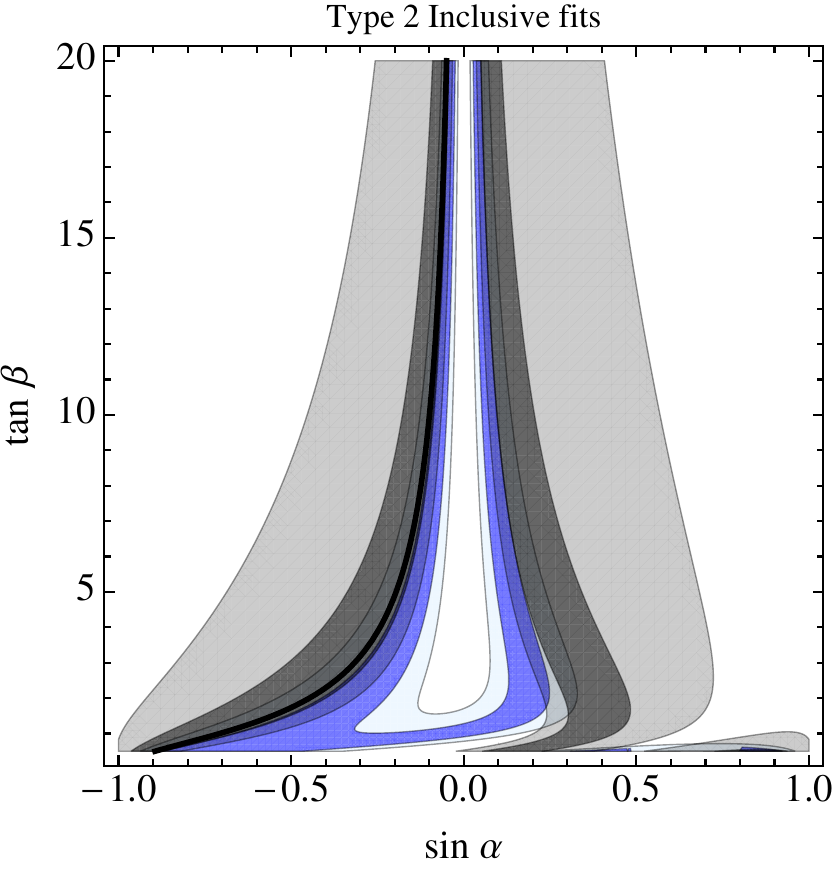} 
 
  \caption{Signal fits to CMS results for 
    a Standard Model-like Higgs boson in 
  Type 1 (left) and Type 2 (right) 2HDM. 
  The dark (light) blue region denotes the $1 \sigma$ ($2 \sigma$) CMS result for 
   inclusive $\sigma \! \cdot \! {\rm Br} (h \to \gamma \gamma) / \sigma \! \cdot \! {\rm Br}(h_{SM} \to \gamma \gamma)$. The dark (light) grey region denotes the $1 \sigma$ ($2 \sigma$) CMS result for 
    inclusive $\sigma \! \cdot \! {\rm Br} (h \to ZZ^*) / \sigma \! \cdot \! {\rm Br}(h_{SM} \to ZZ^*)$. The black line denotes the alignment limit $\sin(\beta - \alpha) = 1$. }
   \label{fig:signalfits}
\end{figure}

In Fig.~\ref{fig:signalfits} we show the $1 \sigma$ and $2 \sigma$ 
allowed ranges in $\sin \alpha, \tan \beta$ for Type 1 and Type 2 2HDM 
consistent with the current CMS results for 
$\sigma \! \cdot \! {\rm Br} (h \to \gamma \gamma) / \sigma \! \cdot \! {\rm Br}(h_{SM} \to \gamma \gamma)$ and $\sigma \! \cdot \! {\rm Br} (h \to ZZ^*) / \sigma \! \cdot \! {\rm Br}(h_{SM} \to ZZ^*)$. 
The ranges for Type 3 and Type 4 2HDM are similar to those of Type 1 and Type 2, respectively. 
The standard deviation ranges are obtained from the quoted CMS combination of statistical and systematic uncertainties. 
As in our earlier discussion of inclusive signals, we approximate the inclusive production $\sigma \! \cdot \! {\rm Br}$ by summing over the production cross sections of various channels with no relative weight. This does not account for possible differences in the experimental acceptance between production channels, which introduces an additional uncertainty of $\mathcal{O}(\sigma^{SM}_{qq' \to h} / \sigma^{SM}_{gg \to h}) \sim 8$\% that does not affect the qualitative result. We add cross sections for 7 TeV and 8 TeV $pp$ collisions weighted by the relative integrated luminosity of each sample in the CMS result (5.1 and 5.3 fb$^{-1}$, respectively). 

These approximate fits to the allowed regions of 2HDM parameter space 
highlight several salient features. 
In Type 1 and Type 3 2HDM, the range of inclusive $\sigma \! \cdot \! {\rm Br} (h \to \gamma \gamma) / \sigma \! \cdot \! {\rm Br}(h_{SM} \to \gamma \gamma)$ signals lies outside the current CMS $1\sigma$ signal fit, due to the fact that CMS observes a substantially enhanced diphoton rate that cannot be accommodated in Type 1 \& Type 3 2HDM.  However, at $2 \sigma$ both the $\sigma \! \cdot \! {\rm Br} (h \to \gamma \gamma) / \sigma \! \cdot \! {\rm Br}(h_{SM} \to \gamma \gamma)$  and   $\sigma \! \cdot \! {\rm Br} (h \to ZZ^*) / \sigma \! \cdot \! {\rm Br}(h_{SM} \to ZZ^*)$ fits are compatible with a wide range of $\alpha, \beta$, since the relevant production cross sections and branching ratios only vary slowly as a function of these mixing angles. As discussed earlier, this is because the contours of the inclusive di-photon and $ZZ$ $\sigma \! \cdot \! {\rm Br}$ largely track the $g_{hVV} = \sin(\beta - \alpha)$ coupling in Type 1 and Type 3 2HDM. Although the branching ratios to photons and $Z$ bosons
increases in inverse proportionality to the total width, the dominant $gg \to h$ 
production rate is directly proportional  to the $h \to b \bar b$ partial width, since the $ggh$ coupling is proportional to the $h t \bar{t}$ coupling and both $g_{ht\bar{t}}, g_{h b \bar{b}} \propto \cos \alpha / \sin \beta$. 
Thus any gains in di-photon or $ZZ$ branching ratios due to decreasing $\Gamma(h \to b \bar b)$ are offset by a decrease in the production rate, and the dominant change in the inclusive $\sigma \! \cdot \! {\rm Br}$ comes from changes in the $hVV$ coupling. 
 The slow variation in these inclusive signals as a function of $\alpha, \beta$ raises the prospect of a substantially enhanced rate for VBF diphoton $\sigma \! \cdot \! {\rm Br}$ in Type 1 \& 3 2HDM consistent with the current measurements, 
 or similarly a substantially suppressed (enhanced) inclusive ditau $\sigma \! \cdot \! {\rm Br}$ for Type 1 (Type 3) 2HDM.

The signal fits are consistent with a much smaller range of $\alpha, \beta$ in Type 2 and Type 4 2HDM, as in these cases the relevant branching ratios vary quickly away from the alignment limit. In contrast with the Type 1 and 3 2HDM, the Type 2 and 4 2HDM 
may accommodate the $1 \sigma$ best fit to 
$\sigma \! \cdot \! {\rm Br} (h \to \gamma \gamma) / \sigma \! \cdot \! {\rm Br}(h_{SM} \to \gamma \gamma)$ 
when $\Gamma(h \to b \bar{b}) / \Gamma(h_{SM} \to b \bar{b}) < 1$, 
though the region of mutual agreement between the $\sigma \! \cdot \! {\rm Br} (h \to \gamma \gamma) / \sigma \! \cdot \! {\rm Br}(h_{SM} \to \gamma \gamma)$  and   $\sigma \! \cdot \! {\rm Br} (h \to ZZ^*) / \sigma \! \cdot \! {\rm Br}(h_{SM} \to ZZ^*)$ is narrow.  As discussed earlier, the rapid variation of these inclusive $\sigma \! \cdot \! {\rm Br}$ in Type 2 and Type 4 2HDM theories arises because the $gg \to h$ production rate and total width are not strongly correlated, so that the diphoton and $ZZ$ branching ratios may increase as the total width drops, while the production rate $\sigma(gg \to h)$ remains fixed. This leads to rapid changes in the inclusive rates as $\Gamma(h\to b \bar b)$ varies with $\alpha$ and $\beta$.

\section{Conclusion}
\label{conclusion}

A variety of inclusive and exclusive standard Higgs search channels should be observable 
for a light SM-like Higgs boson 
with relatively low integrated luminosity at the LHC. 
These standard channels are likely to provide the first indications 
for a SM-like Higgs boson of any deviations of cross section times branching ratios 
from Standard Model expectations.  
In this paper we have focused on the standard channel signatures of 
CP and flavor conserving 
extended electroweak symmetry breaking sectors with two Higgs doublets. 
These models exhibit a variety of novel features which may be used to distinguish various types of 2HDM theories from the Standard Model Higgs and from each other in the event of discrepancies in certain channels.

Several features stand out. The combination of inclusive di-photon, VBF and $Vh$ di-photon, 
and exclusive ditau $\sigma \! \cdot \! {\rm Br}$ is often sufficient to differentiate different 2HDM types from each other and from the Standard Model. If the exclusive di-photon $\sigma \! \cdot \! {\rm Br}$ is significantly enhanced relative to the inclusive rate, this suggests theories in which all quarks couple to the same Higgs doublet. In contrast, if the inclusive and exclusive di-photon $\sigma \! \cdot \! {\rm Br}$ are significantly enhanced this points to theories in which the up-type and down-type quarks couple to separate doublets. These inclusive and exclusive di-photon $\sigma \! \cdot \! {\rm Br}$ may be largely verified by corresponding diboson $\sigma \! \cdot \! {\rm Br}$, while exclusive ditau $\sigma \! \cdot \! {\rm Br}$ resolve the leptonic couplings.

Various ratios of inclusive and exclusive channels may also be observed, enjoying reduced systematics relative to measurements of individual 
$\sigma \! \cdot \! {\rm Br}$. 
The available inclusive ratios are not strongly sensitive to deviations from SM couplings in 2HDMs, in large part because of the parametric similarities in the $hVV$ and $h \gamma \gamma$ couplings. However, exclusive ratios such as the ratio of VBF ditau and VBF diboson $\sigma \! \cdot \! {\rm Br}$ are quite sensitive to deviations from SM couplings, and may vary sharply from the Standard Model prediction.

Intriguingly, the regions of 2HDM parameter space in which the light CP even neutral scalar $h$ exhibits significant deviations from Standard Model Higgs signals also entail significant couplings of the heavier neutral scalars $H$ and $A$ to visible Standard Model states. Perhaps the most promising channels among these are 
$H \to VV$,
and $H,A \to \gamma \gamma$, all of 
 which enjoy considerable reach at the LHC. Notable discrepancies in the Standard Model signals of $h$ 
 can imply that decays of heavier scalars may  be visible in standard Higgs channels, provided the scalars are not too heavy. 

Finally, we note that the potential exclusive signatures of the  MSSM Higgs are a subset of those presented here, as the MSSM is a Type II 2HDM with a specific set of coupling relations that reduce the number of free parameters. In particular, the angle $\alpha$ is fixed in terms of $\tan \beta$ and the Higgs masses. When $\tan \beta > 1$ this typically leads to an enhancement of the $\Gamma(h \to b \bar b)$ partial width, making it difficult to enhance the inclusive and exclusive di-photon $\sigma \! \cdot \! {\rm Br}$ in the MSSM without introducing additional degrees of freedom.


\bigskip
\bigskip

\section*{Acknowledgments}
\noindent
We thank Kfir Blum, Spencer Chang, Jared Evans, Jamison Galloway, and Can Kilic for useful conversations. This work was supported in part by DOE grant DE-FG02-96ER40959.  NC gratefully acknowledges the support of the Institute for Advanced Study.

\appendix

\end{document}